\documentclass[conf]{new-aiaa}
\usepackage[utf8]{inputenc}
\usepackage{subfig}
\usepackage{hyperref}
\usepackage{graphics}
\usepackage{graphicx}
\usepackage{amsmath}
\usepackage[version=4]{mhchem}
\usepackage{siunitx}
\usepackage{longtable,tabularx}
\usepackage{tikz}
\newcommand{\bsu}{\boldsymbol{u}}
\newcommand{\bsq}{\boldsymbol{q}}
\newcommand{\bsf}{\boldsymbol{f}}

\DeclareMathOperator*{\argmax}{arg\max}
\DeclareMathOperator*{\argmin}{arg\min}
\newcommand{\ip}[2]{\left\langle #1, #2\right\rangle}

\usepackage{floatrow}
\usepackage{xcolor}
\definecolor{myOrange}{rgb}{0.8500,0.3250,0.0980}
\definecolor{myYellow}{rgb}{0.9290,0.6940,0.1250}
\definecolor{myGreen}{rgb}{0.4660,0.6740,0.1880}
\definecolor{myBlue}{rgb}{0,0.4470,0.7410}

\graphicspath{{Figures/}}

\begin{document}

\title{
A sparsity-promoting resolvent analysis for the  identification of spatiotemporally-localized amplification mechanisms
}

\author{Barbara Lopez-Doriga\footnote{Graduate Research Assistant, MMAE Department}}
\affil{Illinois Institute of Technology, Chicago, IL 60616}

\author{Eric Ballouz\footnote{Graduate Research Assistant, Graduate Aerospace Laboratories, AIAA Student Member.} and H. Jane Bae \footnote{Assistant Professor, Graduate Aerospace Laboratories, AIAA Member.}}
\affil{California Institute of Technology, Pasadena, CA, 91125}

\author{Scott T. M. Dawson\footnote{Assistant Professor, MMAE Department, AIAA Senior Member.}}
\affil{Illinois Institute of Technology, Chicago, IL 60616}

\graphicspath{{Figures/}}

\maketitle
\begin{abstract}
This work introduces a variant of resolvent analysis that identifies forcing and response modes that are sparse in both space and time. This is achieved through the use of a sparse principal component analysis (PCA) algorithm, which %uses 
formulates the associated optimization problem as a nonlinear eigenproblem that can be solved with an inverse power method. 
We apply this method to parallel shear flows, both in the case where we  assume Fourier modes in time (as in standard resolvent analysis) and obtain spatial localization, and where we allow for temporally-sparse modes through the use of a linearized Navier--Stokes operator discretized in both space and time. 
Appropriate choice of desired mode sparsity allows for the identification of structures corresponding to high amplification that are localized in both space and time. 
We report on the similarities and differences between these structures and those from standard methods of analysis. After validating this space-time resolvent analysis on statistically-stationary channel flow, we next %apply 
implement the methodology %to
on a time-periodic Stokes boundary layer, demonstrating the applicability of the approach %methodology 
to non-statistically-stationary systems.

%and discuss the potential for the method to be extended to study amplification in temporally-evolving flows.

%We apply this method to simple parallel shear flows, demonstrating that we can identify time-localized modes, 
%. We first demonstrating that it can reproduce the 

%We apply this method to simple parallel shear flows, demonstrating that we can identify time-localized modes.

%demonstrating that we can identify time-localized modes.

%We discuss the sim

%This sparse PCA method is applied to a linearized Navier--Stokes operator discretized in both space and time.

%The results of this ana

%In particular, these modes 

\end{abstract}

\section*{Nomenclature}
{\renewcommand\arraystretch{1.0}
\noindent\begin{longtable*}{@{}l @{\quad=\quad} l@{}}
$x$ & streamwise coordinate\\
$y,z$ & wall-normal and spanwise coordinates\\
$h$ & channel half-height\\
$c $ & wave speed \\% in outer and inner units\\
$U_0$ & characteristic (maximum) streamwise velocity of base flow\\
$u_\tau$ & friction velocity, $\sqrt{\tau_w/\rho}$\\
%$A$ & duct aspect ratio, and normalized channel width \\
$k_x$ & wavenumber in the streamwise direction \\
$k_z$ & wavenumber in the spanwise direction \\
$\lambda_x$ & wavelength in the streamwise direction \\
$\lambda_z$ & wavelength in the spanwise direction \\
$\mathcal{L}$ & linearized equations \\
$\omega$ &  temporal frequency \\
$\mathcal{H}_\omega$ & resolvent operator at temporal frequency $\omega$ \\
$\mathcal{H}_t$ & space-time resolvent operator \\
$\sigma$ & singular value, resolvent gain\\
$\bsq$ & state vector \\
$\bsf$ & forcing vector \\
$\bsu$ & velocity field $(u,v,w)$ \\
%$\bsu'$ & fluctuating field \\
%$\bsU$ & mean velocity field $(U,V,W)$ \\
$p$ &  pressure \\
%$u$ & streamwise velocity component \\
%$v,w$ & wall-normal velocity components \\
$\alpha$ & sparsity parameter, weight of $l_1$ norm  \\
$\gamma$ &sparsity parameter, fraction of nonzero terms  \\
$\delta_\Omega$  & Stokes boundary layer thickness, $\sqrt{2\nu/\Omega}$ \\
$\rho$ &  fluid density  \\
$\nu$ &  fluid kinematic viscosity  \\
$\tau_w$ & Wall shear stress \\
$\eta$ & wall-normal vorticity  \\
$\psi$ & resolvent response mode  \\
$\phi$ & resolvent forcing mode \\
$\Omega $ & Stokes boundary layer frequency \\
$\mathcal{R} $ & Real component \\
%$\Lambda_\epsilon$ & $\epsilon$-pseudospectrum \\
$Re$ & (Outer) Reynolds number, $h U_0/\nu$ \\
$Re_\tau$ & friction Reynolds number, $h u_\tau/\nu$ \\
$Re_{\Omega}$ & Stokes boundary layer Reynolds number, $U_0\delta_\Omega/\nu$ \\
$(\cdot^+)$ & Viscous (inner) units
\end{longtable*}}

\section{Introduction}

The use of Fourier transforms in time, and in directions of spatial homogeneity, are ubiquitous across a range of analysis methods in fluid mechanics. 
 For example, in the context of canonical wall-bounded parallel shear flows, asymptotic linear stability and transient growth analysis typically make use of Fourier decomposition in the streamwise and spanwise directions \cite{Schmid_Henningson}, while resolvent analysis \cite{mckeon2010resolvent} additionally employs a Fourier transform in time. 
This use of Fourier decomposition in directions of homogeneity can be well motivated by showing that Fourier modes naturally arise as the outputs of such analyses.
Data-driven analyses such as the proper orthogonal decomposition (POD) also converge to Fourier modes in direction of spatial homogeneity \cite{holmes2012pod}, with temporal Fourier decomposition also emerging when considering the spectral version of POD \cite{lumley1967,towne2018spectral}.
 
One limitation of such analysis methods is that they can be inefficient or ill-equipped to study structures and mechanisms that are highly localized in space and/or time. Data driven methods have been recently formulated to identify localized structures, using either data-driven wavelet-based decomposition \cite{ren2021image,floryan2021discovering}, or conditional \cite{schmidt2019conditional} or windowed \cite{frame2022space} space-time POD. However, there has been limited work modifying equation-based decomposition methods for localized analysis. Note that the time-evolution of spatially localized disturbances have been studied in the context of instability and transition \cite{henningson1993mechanism}.

The natural emergence of Fourier decompositions in both equation-based and data-driven decompositions can often be related to the use of inner products (or equivalently, $l_2$ energy norms) when formulating such methods as optimization problems. It is possible, however, to modify optimization problems such that structures that are localized (sparse) in space and/or time are identified instead of Fourier modes. This can be achieved through the addition of an appropriate $l_1$ norm terms in the relevant optimization problem. This utilizes theory developed in the context of compressed sensing \cite{candes2008introduction}, which allows for sparsity-promoting methods to be solved using convex methods. Such sparsity-promoting ideas have been utilized across a wide range of applications in the past decade, which in a fluids context include the identification of sparse reduced-order models \cite{brunton2016sindy,loiseau2018constrained,rubini2020l1}, identification of a sparse set of active dynamic modes that best represent time-resolved data \cite{jovanovic2014dmdsp}, and in the reconstruction of spectral content from temporally-underresolved data \cite{tu2014compressed}.
In the context of resolvent analysis, recent work by Skene et al.~\cite{skene2022sparseForcing} implements gradient-based Riemannian optimization of $l_1$-based objective functions to identify spatially-sparse forcing modes. The localized structure of the identified modes potentially makes them more useful for practical flow control purposes than standard
resolvent forcing and response modes.

%localized forcing modes have been identified in the work described in Ref.~\cite{skene2022sparseForcing}, via a Riemann optimization algorithm that incorporates an $l_1$ term while enforcing a unit norm on these modes in order to preserve the $l_2$ sense of the gains. 

The present work formulates a variant of resolvent analysis that promotes sparse and localized modes (rather than Fourier modes) in both space and time. Resolvent analysis has been successful in modeling a variety of phenomena emergent in turbulent flows (e.g.~\cite{mckeon2010resolvent,sharma2013resolvent,luhar2015compliant,mckeon2017engine,dawson2019shape,lesshafft2019resolvent,abreu2020spectral,towne2020resolvent,pickering2021resolvent,yeh2020resolvent,bae2020resolvent}). The variant developed here is intended to make resolvent-based methods more amenable for modeling dynamics and processes that are localized in space and time. While we consider statistically stationary-in-time flows in the present work, we are ultimately motivated by a desire to extend such analysis to temporally-evolving systems. 
The recently-developed Harmonic resolvent analysis \cite{harmonic2020padovan,padovan2022analysis} allows for the study of the amplification properties of systems with periodic base/mean flows.

%This prospect draws some connections to the cross-frequency analysis described in Ref.~\cite{harmonic2020padovan,padovan2022analysis} for a steady base flow.

The structure of the paper is as follows. In Sec.~\ref{sec:method}, we formulate this sparse resolvent analysis, before presenting results applying it to parallel shear flows in Sec.~\ref{sec:results}. In Sec.~\ref{sec:resultsChannel}, we seek and identify modes that are statistically-stationary and sparse in the wall-normal direction, sparse in the wall-normal and spanwise directions, or sparse in the wall-normal and time dimensions. In Sec.~\ref{sec:resultsStokesBL}, we discuss the implementation of this formulation  a Stokes boundary layer, 
 which is a non-stationary time-periodic system for which traditional resolvent analysis is not applicable. %In this case, the observed space-time resolvent modes  may be periodic (butin the time dimension and/or sparse the wall-normal and time dimensions. 

\section{Methodology}
\label{sec:method}
Here, we briefly describe the different resolvent formulations adopted in this study:  Sec.~\ref{sec:regularResolvent} introduces traditional resolvent analysis, which assumes that the systems are statistically-stationary in the time dimension and homogeneous in the streamwise and spanwise spatial directions (i.e.~assuming Fourier modes for these spatial structures); Sec.~\ref{sec:spaceTimeResolvent} considers homogeneity only in the space dimension and includes a time differential operator; and Sec.~\ref{sec:sparseResolvent} presents a formulation of resolvent analysis that enforces sparsity on the resolvent modes in both temporal and spatial dimensions. 

\subsection{Resolvent analysis}
\label{sec:regularResolvent}

We start by considering a dynamical system of the form
\begin{equation} % removed t dependence of A
    \dot{\boldsymbol{x}} + \boldsymbol{A}  \boldsymbol{x} =\boldsymbol{f},
    \label{eq:dynSys}
\end{equation}
with $\boldsymbol{x}$, $\boldsymbol{f} \in \mathbb{R}^n$, and $\boldsymbol{A} \in \mathbb{R}^{n \times n}$. Here, $\boldsymbol{A}$ is a linear operator, $\boldsymbol{x}$ represents the  state of the system, and the term $\boldsymbol{f}$ represents an external input, which could come from neglected nonlinear terms.  
In this subsection, we proceed by taking a Fourier transform in time, thus considering solutions for both the state of the system and the forcing terms that are of the form.
\begin{equation}
    \boldsymbol{x} = \hat{\boldsymbol{x}} \exp{(-i \omega t)},
    \label{eq:FourierX}
\end{equation}
\begin{equation}
    \boldsymbol{f} = \hat{\boldsymbol{f}} \exp{(-i \omega t)},
    \label{eq:FourierF}
\end{equation}
with $\omega \in \mathbb{C}$ denoting the temporal frequency.  
Substituting Eqs.~(\ref{eq:FourierX})-(\ref{eq:FourierF}) into Eq.~(\ref{eq:dynSys}) gives
\begin{equation}
    (-i\omega \boldsymbol{I}+\boldsymbol{A})\hat{\boldsymbol{x}} = \hat{\boldsymbol{f}}.
\end{equation}
In the case where $i \omega$ is not in the spectrum of the operator $\boldsymbol{A}$, this equation can be recast as
\begin{equation}
    \hat{\boldsymbol{x}} = (-i\omega \boldsymbol{I}+\boldsymbol{A})^{-1}\hat{\boldsymbol{f}}:= \mathcal{H}_\omega \hat{\boldsymbol{f}},
    \label{eq:resolvent}
\end{equation}
for which an arbitrary external forcing $\hat{\boldsymbol{f}}$ is mapped to a given state $\hat{\boldsymbol{x}}$ via the resolvent operator $\mathcal{H}_\omega$, defined at a given frequency, $\omega$. 

In this work, we are interested in identifying cases where a forcing of a small magnitude $\hat{\boldsymbol{f}}$ produces a greatly amplified response $\hat{\boldsymbol{x}}$. This can be achieved through a pseudospectral analysis of the resolvent operator $\mathcal{H}_\omega$, through the singular value decomposition (SVD)
\begin{equation}
    \mathcal{H}_\omega = \sum_{j=1}^N \psi_j \sigma_j \phi_j^*,
    \label{eq:SVD}
\end{equation}
where $(\cdot^*)$ denotes the adjoint.
%provides a more efficient basis for these spatial structures.
In this decomposition, the singular values are sorted by decreasing energy content, such that $\sigma_k \geq \sigma_{k+1} \geq 0$ for all $k$. Additionally, the leading singular value and vectors are solutions to the following optimization problems
\begin{align}
    \sigma_1 &=\max_{\|\phi\| = 1}\|\mathcal{H}_\omega \phi\| = \max_{\|\psi\| = 1}\|\mathcal{H}^*_\omega \psi\|, \\
%\end{equation}
%and
%\begin{equation}
    \phi_1&=\argmax_{\|\phi\|=1} ||\mathcal{H}_\omega \phi||, \\
%\end{equation}
%with 
%\begin{equation}
    \psi_1& =\sigma_1^{-1} \mathcal{H}_\omega \phi_1 =\argmax_{\|\psi\|=1} ||\mathcal{H}^*_\omega \psi||.
\end{align}
Here, we formulate the resolvent operator for the nondimensionalized incompressible Navier-Stokes equations
\begin{equation}
\label{eq:momentum}
\frac{\partial \bsu}{\partial t}+(\bsu \cdot \nabla) \bsu = -\nabla p+\frac{1}{Re}\Delta \bsu,
\end{equation}
\begin{equation}
\label{eq:continuity}
\nabla \cdot \bsu =0,
\end{equation}
where $\bsu=(u,v,w)$ is the velocity field, and $p$ is the pressure. When laminar base flows are considered, these equations are nondimensionalized based on the channel half-height, $h$, and the maximum streamwise flow speed, $U_0$. The Reynolds number is thus  $Re = h U_0/\nu$, where $\nu$ is the kinematic viscosity. For turbulent flow, velocity is instead nondimensionalized using the friction velocity $u_\tau = \sqrt{\tau_w/\rho}$, with $\tau_w$ denoting the wall shear stress, and $\rho$ the fluid density. In the turbulent case, we thus use the friction Reynolds number $Re_\tau = h u_\tau/\nu$ in Eq.(~\ref{eq:momentum}). 
This system of equations is expressed in the form of Eqs.~(\ref{eq:dynSys})-(\ref{eq:resolvent}) by linearizing about either a laminar equilibrium or turbulent mean flow.

In this study we consider wall-bounded parallel flows with a unidirectional mean/base flow. This makes it convenient to formulate Eq.~(\ref{eq:resolvent}) in terms of the wall-normal velocity, $v$, and vorticity, $\eta = \partial u/\partial z-\partial w/\partial x$, Assuming Fourier transforms in the streamwise and spanwise directions with wavenumbers $k_x$ and $k_z$ respectively, this gives
%that is, a system with a only a nonzero velocity component $\textbf{u}=(u(y),0,0)$ in the $x$-direction. Hence, instead of using a formulation of the governing equations that includes the streamwise velocity $u$ and pressure $p$, we adopt the wall-normal velocity ($v$) and vorticity ($\eta=\partial u/\partial z-\partial w/\partial x$) components. %Now, as was introduced in the beginning of this subsection, here we will be assuming that the properties of the system are statistically stationary in time and homogeneous in the streamwise direction, allowing 
%We consider trajectories in the form of Eq.~\ref{eq:FourierF} for both $v'$ and $\eta'$ such that
% \begin{equation}
% \label{eq:vdecomp}
% v' (\bsx,t) = \hat{v}(y) \exp{[{i}(k_x x+k_z z-\omega t)]},
% \end{equation}
% \begin{equation}
% \label{eq:etadecomp}
% \eta' (\bsx,t) = \hat{\eta}(y) \exp{[{i}(k_x x+k_z z-\omega t)]},
% \end{equation}
% where $k_x$ and $k_z$ represent the streamwise and spanwise wavenumbers, respectively, and $\hat{\eta} = ik_z\hat{u}-ik_x\hat{w}$. Substituting Eqs.~\ref{eq:vdecomp}-\eqref{eq:etadecomp} into Eqs.~\ref{eq:fluct}-\eqref{eq:continuity_fluct} gives the resolvent operator in terms of the formulation introduced in Eq.~\ref{eq:resolvent}
\begin{equation}
\label{eq:wallNormalResolvent}
\begin{pmatrix}
\hat{v} \\
\hat{\eta}
\end{pmatrix}=
\begin{pmatrix}
-i\omega + \Delta^{-1}\mathcal{L}_{os} & 0 \\
i k_z U_y & -i\omega+\mathcal{L}_{sq}
\end{pmatrix}^{-1}
\begin{pmatrix}
\hat{f}_v \\
\hat{f}_\eta
\end{pmatrix}=\mathcal{H}_\omega  
\begin{pmatrix}
\hat{f}_v \\
\hat{f}_\eta
\end{pmatrix},
\end{equation}
where $U_y$ represents the gradient of the mean streamwise velocity along the wall-normal dimension $y$, and $\Delta = \partial_{yy}-(k_x^2+k_z^2)$ is the Laplacian operator. %Note how all the nonlinear terms are condensed in a single forcing term $\textbf{f}$, placed in the right-hand side. 
The Orr-Sommerfeld (OS) and Squire (SQ) operators are 
\begin{equation}
\label{eq:OSop}
    \mathcal{L}_{os}=i k_x U\Delta-i k_x U_{yy}-\frac{1}{Re}\Delta^2,
\end{equation}
\begin{equation}
\label{eq:SQop}
    \mathcal{L}_{sq}=i k_x U-\frac{1}{Re} \Delta,
\end{equation}
where $U_{yy}$ represents the second derivative of the mean streamwise velocity in the wall-normal direction. Note that the Fourier-transformed quantities $\hat v$ and $\hat \eta$ for a given set of wavenumbers can be expressed in  physical (mean subtracted) variables by
\begin{equation}
\label{eq:vdecomp}
v(x,y,z,t) = \hat{v}(y) \exp{[{i}(k_x x+k_z z-\omega t)]},
\end{equation}
\begin{equation}
\label{eq:etadecomp}
\eta (x,y,z,t) = \hat{\eta}(y) \exp{[{i}(k_x x+k_z z-\omega t)]}.
\end{equation}

% PUT IN CASE WITH NO FOURIER TRANSFORM IN Z?

\subsection{Space-time resolvent analysis}
\label{sec:spaceTimeResolvent}
We now consider the case without a Fourier transform in the temporal direction. While we restrict our attention to statistically-stationary flows in the present work, this formulation will ultimately be applicable for flows where the ensemble-averaged state evolves in time. % time-evolving
Much of the description described in Sec.~\ref{sec:regularResolvent} can be similarly applied, without a Fourier transform in time. In particular, Eq.~(\ref{eq:dynSys}) becomes
%In this analysis we will not assume periodicity in the temporal dimension. Instead, we can rearrange the terms in Eq.~\ref{eq:dynSys} such that
\begin{equation}
\label{eq:dynSysT}
    (D_t+\boldsymbol{A}(t)) \boldsymbol{x} = \boldsymbol{f},
\end{equation}
where we introduce $D_t$ as a time differentiation operator. Similarly, the equivalent of Eq.~(\ref{eq:wallNormalResolvent}) is
\begin{equation}
\label{eq:wallNormalResolventwithT}
\begin{pmatrix}
\hat{v} \\
\hat{\eta}
\end{pmatrix}=
\begin{pmatrix}
\frac{\partial}{\partial t} + \Delta^{-1}\mathcal{L}_{os} & 0 \\
i k_z U_y & \frac{\partial}{\partial t}+\mathcal{L}_{sq}
\end{pmatrix}^{-1}
\begin{pmatrix}
\hat{f}_v \\
\hat{f}_\eta
\end{pmatrix}=\mathcal{H}_t 
\begin{pmatrix}
\hat{f}_v \\
\hat{f}_\eta
\end{pmatrix}.
\end{equation}
Note that here the $\hat\cdot$ notation now refers to a Fourier transform in the $x$ and $z-$directions only, and that $\hat v$ and $\hat \eta$ now have explicit time dependence, with the equivalent of Eqs.~(\ref{eq:vdecomp})-(\ref{eq:etadecomp}) being
\begin{equation}
\label{eq:vdecompT}
v(x,y,z,t) = \hat{v}(y,t) \exp{[{i}(k_x x+k_z z)]},
\end{equation}
\begin{equation}
\label{eq:etadecompT}
\eta(x,y,z,t) = \hat{\eta}(y,t) \exp{[{i}(k_x x+k_z z)]}.
\end{equation}
 Note that the  resolvent operator described in Eq.~(\ref{eq:wallNormalResolventwithT}) will be of a much larger dimension when discretized than Eq.~(\ref{eq:wallNormalResolvent}), as time is no longer decoupled. 
  We refer to this formulation as \emph{space-time resolvent analysis}, though we note that not taking a Fourier transform in time reduces the direct connection with the resolvent operator associated with a linear dynamical system. 
 For a system with constant (in time) mean, a singular value decomposition of the space-time resolvent operator in Eq.~(\ref{eq:wallNormalResolventwithT}) should identify the same forcing and response modes as those from  Eq.~(\ref{eq:wallNormalResolvent}) at each frequency that can be captured by discretization. The following section will describe a variant of resolvent analysis for which this is no longer the case.

\subsection{Sparse resolvent analysis}
\label{sec:sparseResolvent}

To formulate a variant of resolvent analysis that promotes sparsity in resolvent modes, we utilize a version of sparse principal component analysis. The approach we follow is described in Ref.~\cite{hein2010inverse}. Similar methods are also described and discussed in Refs.~\cite{jolliffe2003modified,zou2006sparse,sigg2008expectation,journee2010generalized,zou2018selective}. 

Note that we may seek sparsity in either the forcing or response modes. In general, we find that for the problems considered in this work, similar results are typically obtained in both cases. 
We formulate the problem here assuming that sparsity is desired in the response modes, $\psi_j$. First, we note that the leading resolvent response mode, $\psi_1$ satisfies
\begin{equation}
    \psi_1 = \argmax_\psi \frac{\ip{\psi} {\mathcal{H}\mathcal{H}^*\psi}}{\|\psi\|_2^2 } = \argmax_\psi \frac{\|\mathcal{H}^*\psi\|^2_2}{\|\psi\|_2^2} = \argmax_\psi \frac{\|\mathcal{H}^*\psi\|_2}{\|\psi\|_2}.
\end{equation}
By inverting these expressions, we equivalently have 
\begin{equation}
\label{eq:minsvd}
    \psi_1 = %\argmin_\psi \frac{\|\psi\|_2^2 }{\ip{\psi} {\mathcal{H}\mathcal{H}^*\psi}} 
     \argmin_\psi \frac{\|\psi\|_2}{\|\mathcal{H}^*\psi\|_2}.
\end{equation}
To promote sparsity in  $\psi_1$, we replace the numerator on the right hand side of Eq.~(\ref{eq:minsvd}) with a convex function that includes contributions from both the $l_1$ and $l_2$ norms of $\psi$, replacing the fraction, giving the sparsity-promoting variant of Eq.~(\ref{eq:minsvd}) as
\begin{equation}
\label{eq:minsvdsparse}
    \psi_1 =  \argmin_\psi \frac{(1-\alpha)\|\psi\|_2+ \alpha\|\psi\|_1 }{\|\mathcal{H}^*\psi\|_2}.
\end{equation}
Here $\alpha \in [0,1]$ is a parameter than controls the sparsity of $\psi$, where standard resolvent analysis is recovered with $\alpha = 0$, and $\alpha =1$ corresponds to the sparsest nontrivial solution obtainable by this method. Rather than referring to this parameter explicitly, we will typically refer to a related parameter, $\gamma$, denoting the relative sparsity of the identified sparse modes
\begin{equation}
    \gamma(\psi) = \frac{\|\psi\|_0}{\text{length}(\psi)},
\end{equation}
where $\|\psi\|_0$ denotes the $l_0$ (pseudo)norm of $\psi$ (the number of nonzero entries), and $\text{length}(\psi)$ denotes the total number of entries in $\psi$. 
The optimization problem given in Eq.~(\ref{eq:minsvdsparse}) is solved by formulating a nonlinear eigenvalue problem that can be solved using using an inverse power method, according to the methodology described in Ref.~\cite{hein2010inverse}. % and presented as follows

In practice, we find that this method identifies modes that quite abruptly jump from nonzero to zero values. To counter this, we may pass identified modes through the resolvent operator to obtain more physically-relevant (yet still sparse) modes. That is, we follow the following steps to perform sparse resolvent analysis:
\begin{enumerate}
 \item Compute sparse response modes $\psi_1$ by solving Eq.~(\ref{eq:minsvdsparse})
 \item Compute corresponding forcing modes 
 \begin{equation}
 \label{eq:forcing}
     \phi_1 = \frac{\mathcal{H}^*\psi_1}{\| \mathcal{H}^*\psi_1\|_2}
 \end{equation}
 \item Compute updated response modes via
 \begin{equation}
 \label{eq:responseupdated}
      \psi_1 = \frac{\mathcal{H}\phi_1}{\| \mathcal{H}\phi_1\|_2}
 \end{equation}
 \item Compute corresponding singular values via
 \begin{equation}
     \sigma_1 = \| \mathcal{H}\phi_1\|_2
 \end{equation}
\end{enumerate}
Note in particular that while the $l_1$ norm is used to promote sparsity, we still normalize the modes and compute the associated gains (singular values) based on the $l_2$ norm. 
To compute additional resolvent modes, a deflation scheme is used to project out the components already identified, as described in Ref.~\cite{buhler2014flexible}. 
We can similarly solve for sparse forcing modes by exchanging $\psi$ and $\phi$, and $\mathcal{H}$ and $\mathcal{H}^*$ in this analysis.
% mention that modes are not orthogonal?

\subsection{Problem setup and numerical methods}
The methodology defined in Secs.~\ref{sec:regularResolvent}-%\ref{sec:spaceTimeResolvent}
\ref{sec:resultsStokesBL}
will be applied to study planar flow between two parallel plates, in several contexts.  %We will consider both pressure-driven flow (either with a parabolic laminar base flow or turbulent mean), and Couette flow driven by the relative motion of the boundaries. 
In Secs.~\ref{sec:regularResolvent}-\ref{sec:spaceTimeResolvent} we first consider  pressure-driven flow, either with a parabolic laminar base flow or turbulent mean. 
The turbulent mean is computed from direct numerical simulations (DNS) at a friction Reynolds number of 186, using code described and validated in previous studies \cite{bae2018turbulence,bae2019dynamic,lozano2019characteristic}. This code utilizes a staggered second-order finite difference scheme \cite{orlandi2000fluid}, with a fractional step method \cite{kim1985application} and third-order Runge-Kutta timestepping \cite{wray1990minimal}. 
Sec.~\ref{sec:resultsStokesBL} considers turbulent Stokes boundary layer flow between two oscillating plates, with the time-periodic mean flow computed from data generated using the same DNS solver.
%When a turbulent mean is used, it is estimated using an eddy viscosity model \cite{reynolds1967stability}.

For resolvent analysis, 
the spatial domain in the wall-normal direction is discretized using a Chebyshev collocation method. When the spanwise and time dimensions are explicitly discretized, a Fourier discretization (with periodic boundary conditions) is used. We use the package described in Ref.~\cite{weideman2000matlab} to form both the Chebyshev and Fourier differentiation operators. The number of collocation points used in the spatial and temporal dimensions varies between the examples considered, and will be noted in each section.

\section{Results}
\label{sec:results}
Here, we present results for 3 different cases.  Sec.~\ref{sec:sparseSpace} considers  spatial resolvent modes obtained via both traditional and sparse resolvent analysis for turbulent channel flow, where we also consider the case where the spanwise direction is explicitly discretized rather than Fourier-transformed.   Sec.~\ref{sec:resultsChannel} considers space-time resolvent analysis of laminar channel flow. We first show that we recover the temporal Fourier transform when applying a standard $l_2$ optimization, before showing that temporally-sparse modes are obtained with a modified ($l_1$-based) optimization. 
Sec.~\ref{sec:resultsStokesBL} applies these standard and sparsity-promoting space-time resolvent analysis to a turbulent Stokes boundary layer, where the mean flow is periodic in time.

%$\&$ 
%~\ref{sec:resultsStokesBL} we first show the spatio-temporal resolvent modes of the operator introduced in Eq.~\ref{eq:wallNormalResolventwithT} and consequently discuss the sparse resolvent modes obtained by sparse PCA of the spatio-temporal resolvent operator for planar Poiseuille flow and a turbulent Stokes boundary layer, respectively.  

% Spanwise Channel
\subsection{Spatially-sparse resolvent analysis of turbulent channel flow}
\label{sec:sparseSpace}
In this section, we apply sparse resolvent analysis while using a standard Fourier transform in time. We  first consider channel flow with Fourier transforms additionally applied in the streamwise and spanwise directions, as is standard for such parallel flows. For this case, we consider turbulent channel flow with friction Reynolds number $Re_\tau= u_\tau h/\nu = 186$, where $u_\tau =\sqrt{\tau_w/\rho}$ is the friction velocity. To start with, we consider a 1D analysis where streamwise and spanwise wavenumbers are chosen to give wavelengths $\lambda_x^+ = 1000$  and $\lambda_z^+ = 100$, which matches the typical size of streamwise streaks and vortices associated with the near-wall cycle \cite{jimenez1999autonomous}.  Here and throughout, the $(\cdot^+)$ superscript denotes viscous (inner) units, where velocities are nondimensionalized by the friction velocity $u_\tau$, and lengths by $\nu/u_\tau$.  The temporal frequency chosen gives a wavespeed in inner units of $c^+ = 14.66$. Here and throughout, time is implicitly nondimensionalized by the maximum streamwise flow speed, $U_0$, and channel half-height, $h$. The wall-normal direction is discretized using 201 Chebyshev collocation points.

\begin{figure}
% from ChannelResolventVelocityVorticitySparseBlocks/testResolventL1.m
 \centering 
 \subfloat[]{\includegraphics[width= 0.45\textwidth]{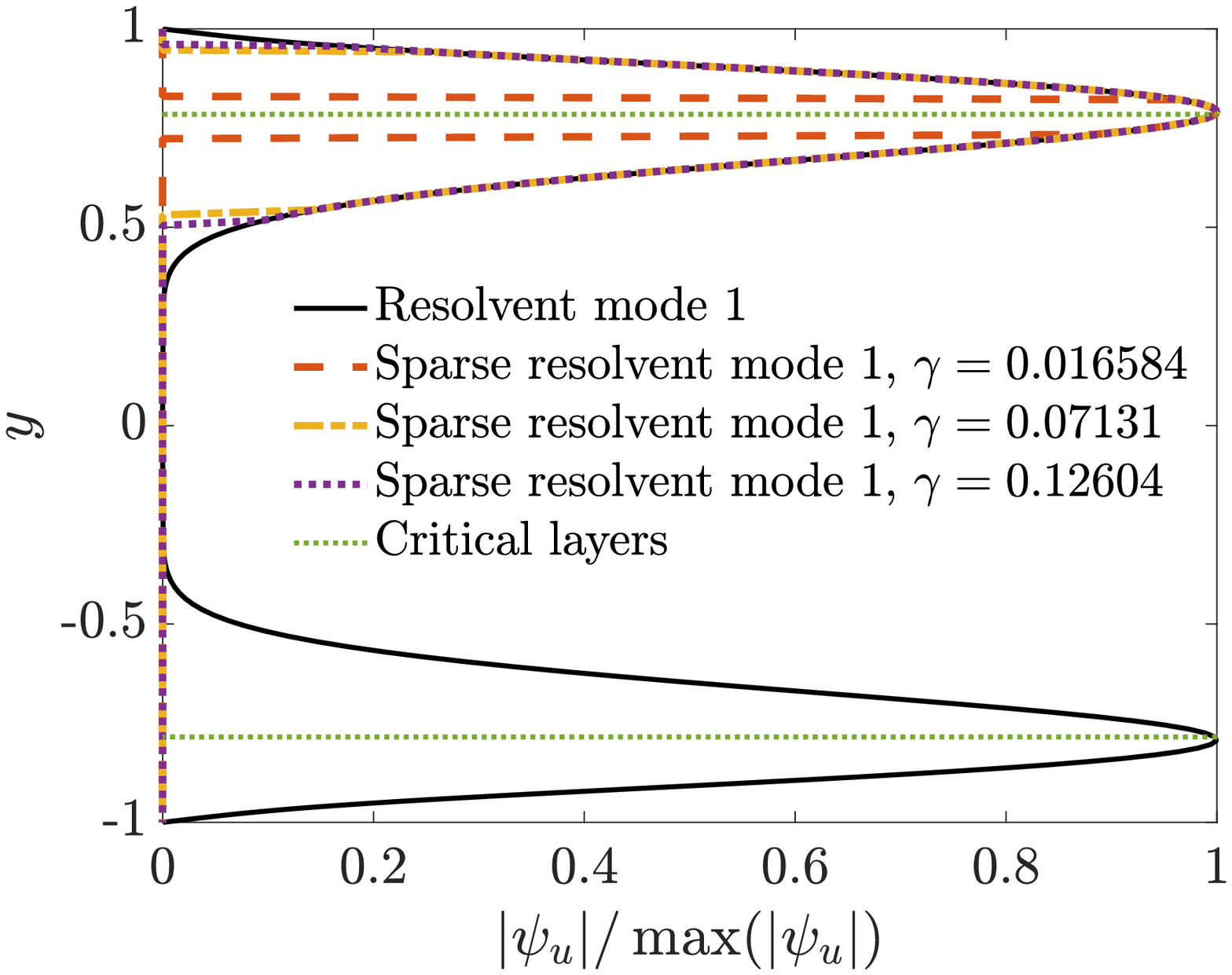}} \ \ 
 \subfloat[]{\includegraphics[width= 0.45\textwidth]{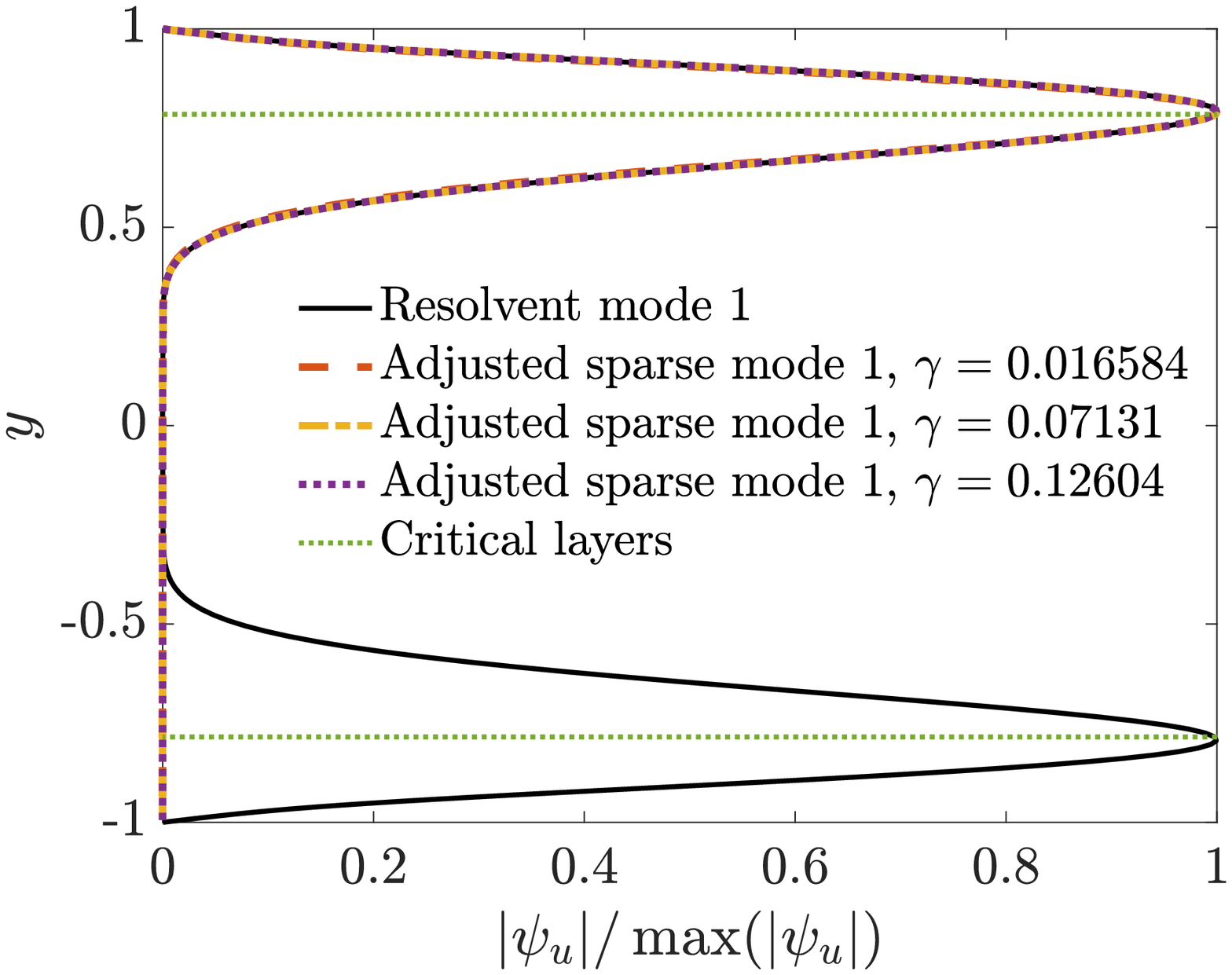}} \\
  \subfloat[]{\includegraphics[width= 0.45\textwidth]{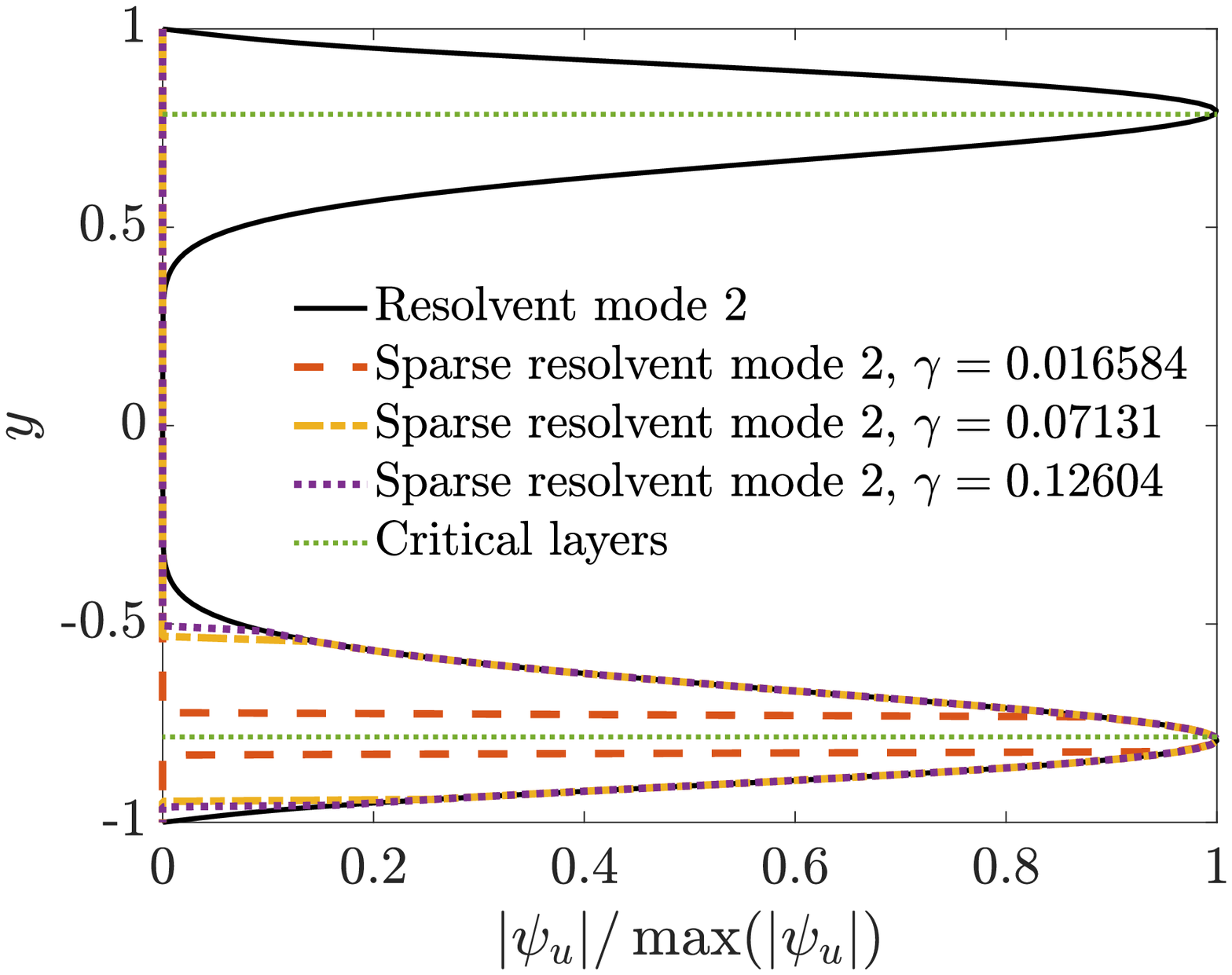}} \ \
 \subfloat[]{\includegraphics[width= 0.45\textwidth]{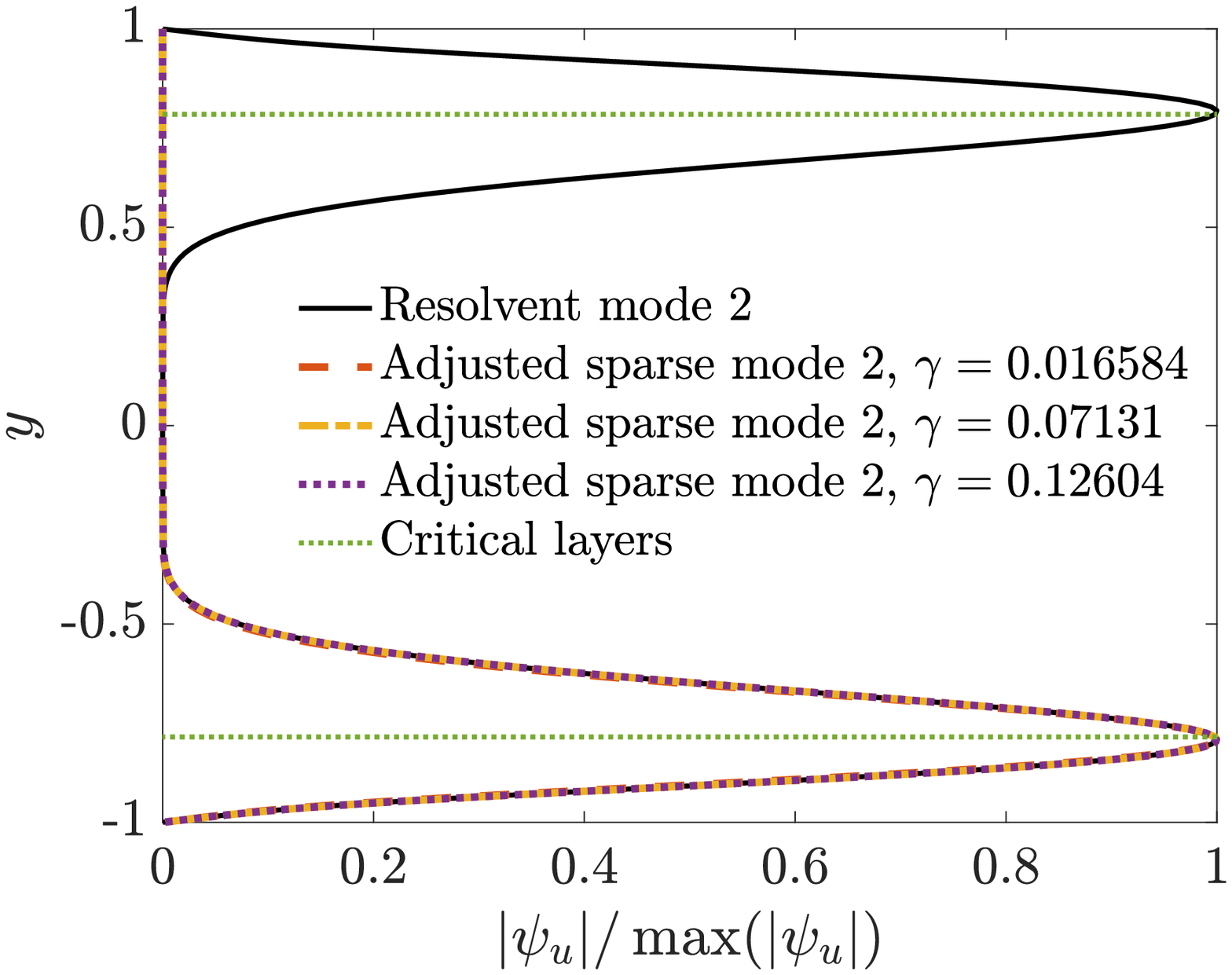}}
\caption{Comparison between leading (a-b) and second (c-d) standard and sparse resolvent response mode amplitudes computed using (a,c) Eq.~(\ref{eq:minsvdsparse}) and (b,d) Eq.~(\ref{eq:responseupdated}). Modes computed for turbulent channel flow with $Re_\tau = 186$, $\lambda^+_x \approx 1000$, $\lambda^+_z = 100$, and $c^+ = 14.66$.
}
\label{fig:1Dmodes}
\end{figure}

Fig.~\ref{fig:1Dmodes} shows the leading two resolvent modes for these parameters, identified using both standard and sparse resolvent analysis (showing only the streamwise velocity components). In this case,  Due to the symmetry of the geometry and mean velocity profile about the centerline ($y=0$), standard resolvent analysis gives leading modes with two peaks, each localized near a critical layer. The first two standard resolvent response modes here have almost identical mode amplitudes, but are orthogonal due to a phase shift between the two peaks. Note that provided that the two peaks are sufficiently separated, the singular values corresponding to the first two modes will also be very similar.  Sparse resolvent analysis, on the other hand, identifies modes that are localized at only one of these two peaks.  Fig.~\ref{fig:1Dmodes}(a) and (c) show the first two sparse resolvent modes identified from Eq.~(\ref{eq:minsvdsparse}) with various values of the sparsity parameter, $\gamma$. While these sparse modes are dependent on  the choice of $\gamma$ (through the choice of $\alpha$ in Eq.~(\ref{eq:minsvdsparse}), it is observed  in Fig.~\ref{fig:1Dmodes}(b) and (d)  that this dependence is substantially reduced when adjusting the modes to be more physically realistic with Eq.~(\ref{eq:responseupdated}). Indeed, for all choices of $\gamma$ considered here, we find that each of the sparse resolvent modes identifies one of the two peaks present in the first two standard resolvent modes. Furthermore, the first two sparse resolvent modes identified in this manner give a subspace very similar to that obtained using the first two standard resolvent modes. In other words, in this case each sparse resolvent mode shown in Fig.~\ref{fig:1Dmodes}(b) and (d) can be closely approximated using a linear combination of the first two standard resolvent modes. In this case, the singular values for the first two sparse resolvent modes are very similar to those for standard resolvent analysis. 
While these results are perhaps unsurprising, this example demonstrates that the sparse resolvent analysis method is behaving as expected.

\begin{figure}
% from Resolvent2D3C_channel/quick_example_sparse
\centering 
  \subfloat[]{\includegraphics[width= 0.5\textwidth]{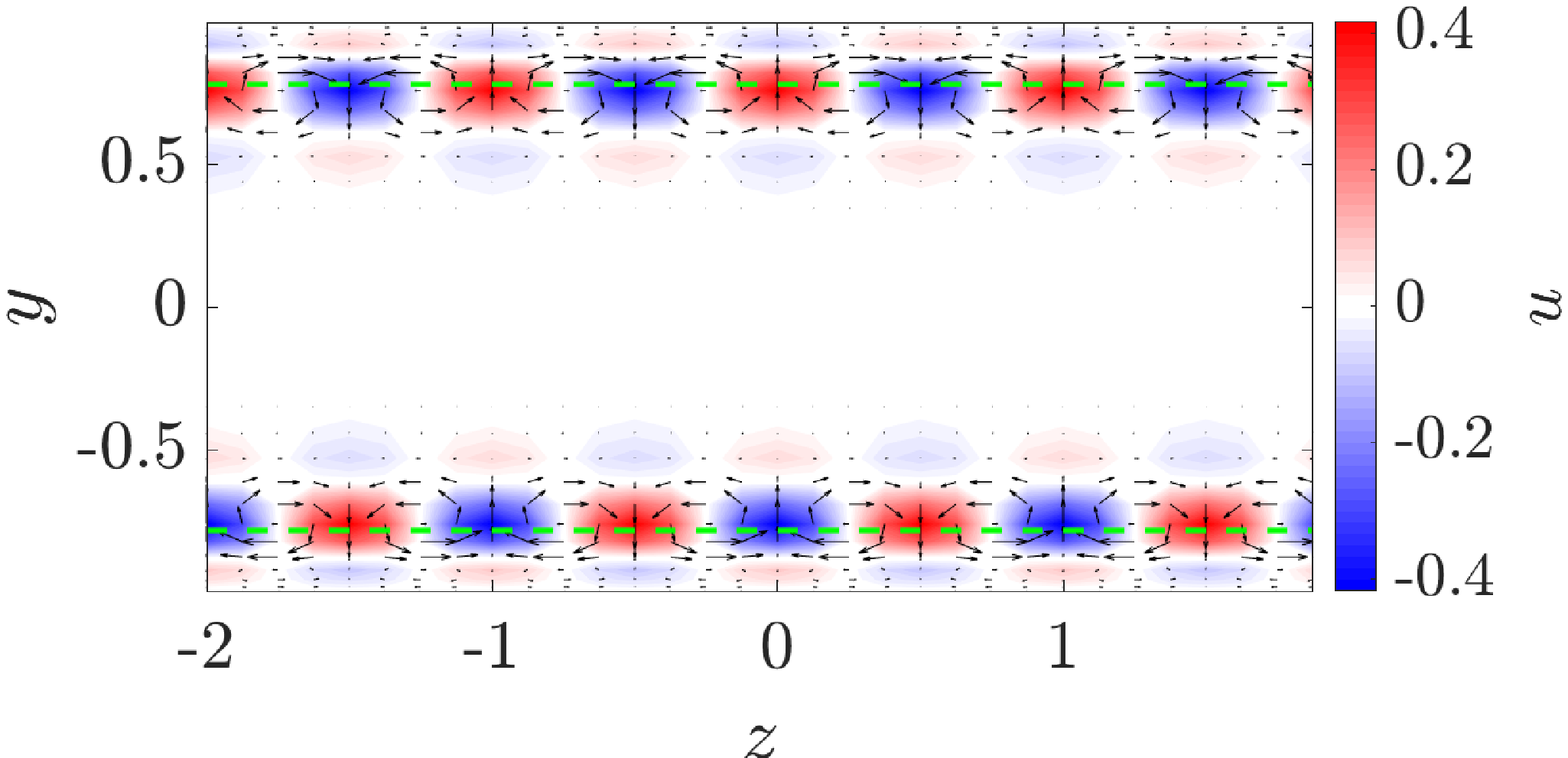}} 
 \subfloat[]{\includegraphics[width= 0.5\textwidth]{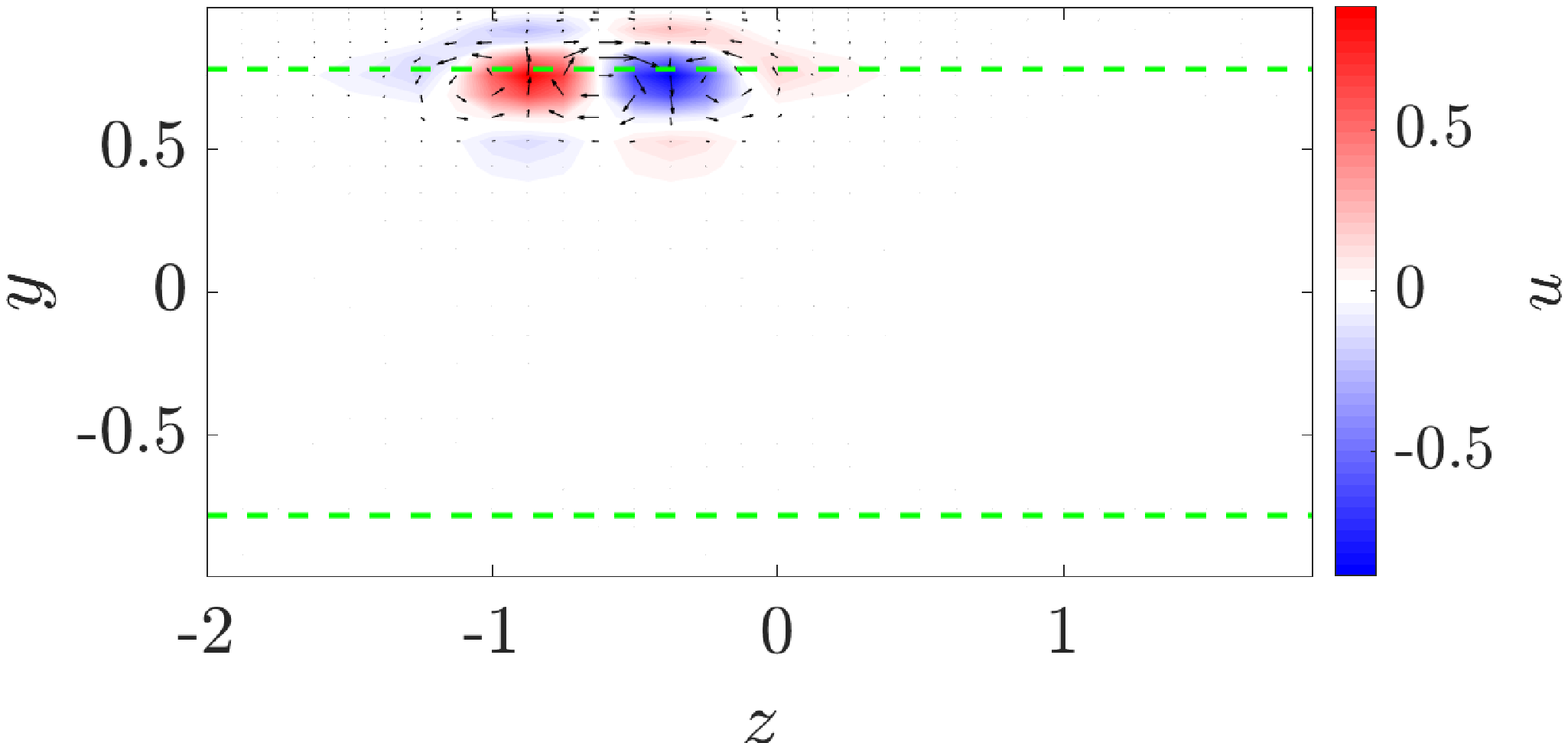}}  \\
 \subfloat[]{\includegraphics[width= 0.5\textwidth]{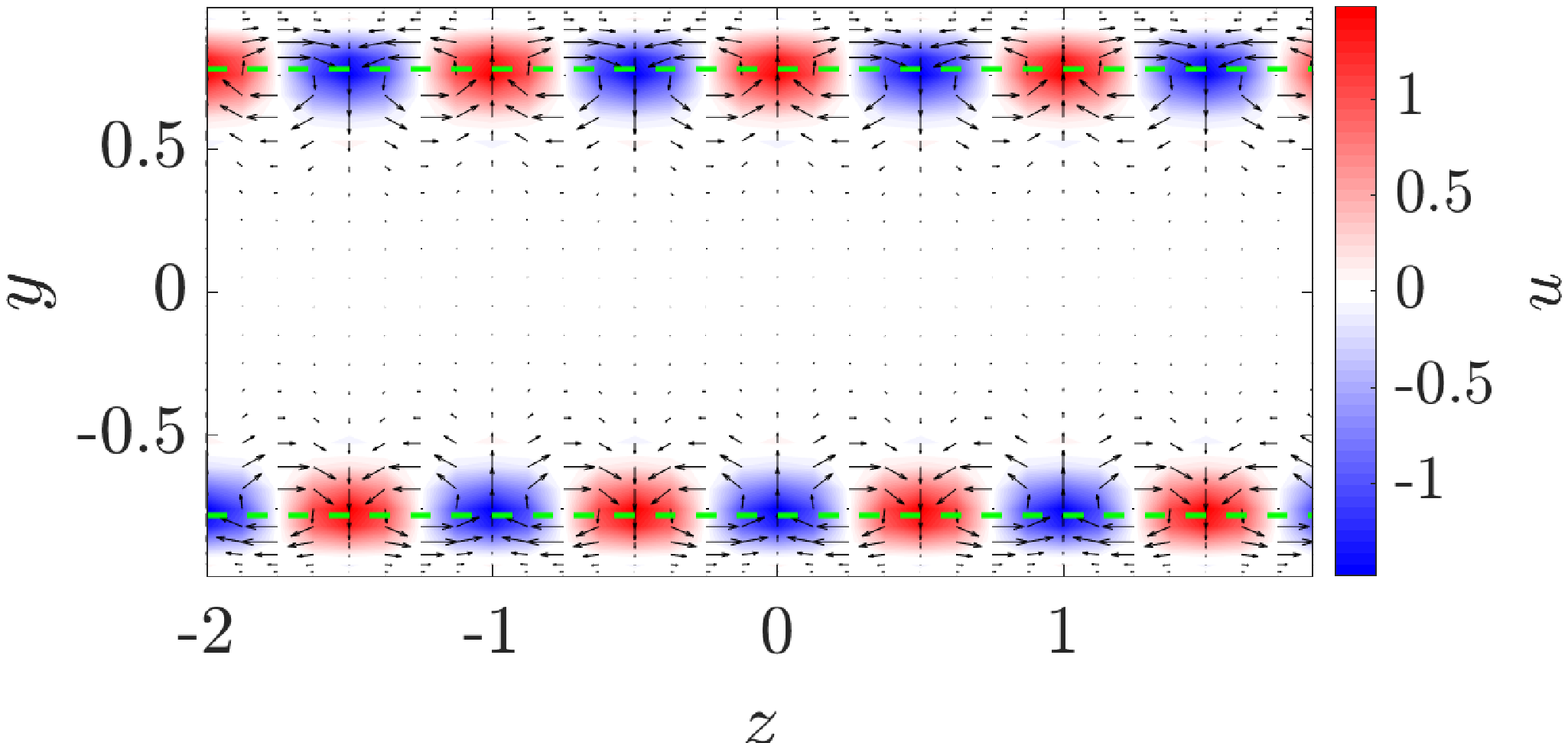}} 
 \subfloat[]{\includegraphics[width= 0.5\textwidth]{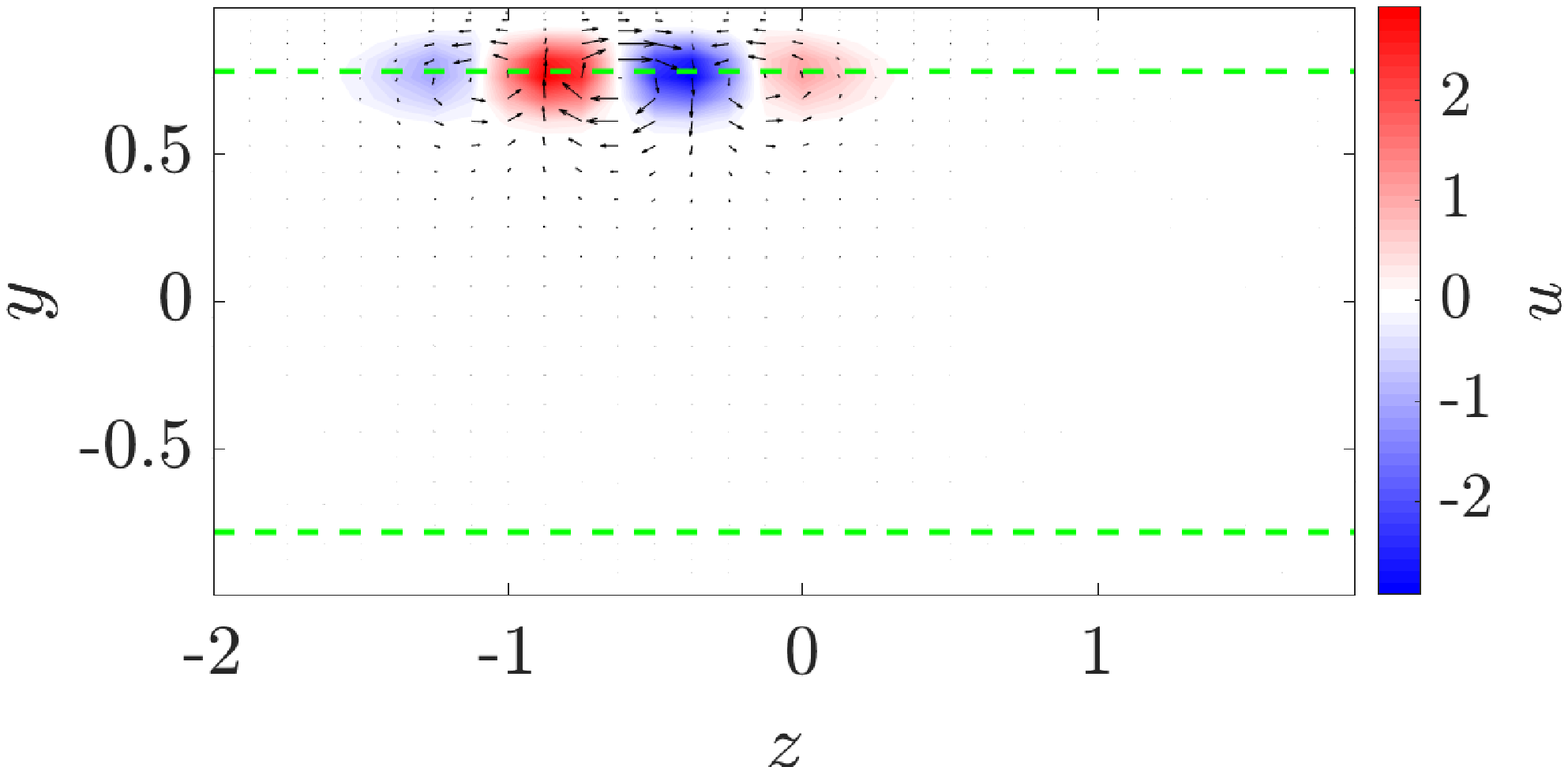}} 
\caption{Comparison between leading resolvent modes for (a) standard and (b) sparse resolvent analysis, applied to turbulent channel flow at $Re_\tau = 
 186$, with a periodic domain in the spanwise direction. Contours of streamwise velocity are shown, with arrows indicating the velocity in the spanwise ($z$) and wall-normal ($y$) directions. Green dashed lines indicate critical layer locations. The streamwise wavelength is $\lambda^+$ = 1000, and wavespeed $c^+=14.66$.}
\label{fig:2Dmodes}
\end{figure}

We next consider a case with the same parameters, but with a periodic domain in the spanwise direction, which we explicitly discretize rather than taking a Fourier transform.  The numerical domain $[-h,h]\times[-2h,2h]$ is discretized using 32 Chebyshev and Fourier modes in the wall-normal and spanwise directions, respectively. Fig.~\ref{fig:2Dmodes} shows representative leading resolvent forcing and response from applying both standard and sparse resolvent analysis. Standard resolvent analysis gives modes that span the entire spanwise extent, consisting of alternating streaks of fast- and slow-moving fluid in the streamwise direction, located near the critical layers. These correlate with wall-normal velocity towards and away from the wall, respectively, indicative the presence of a lift-up amplification mechanism \cite{landahl1975wave,landahl1980note}. The wall-normal and spanwise velocity components form streamwise vortices located between streamwise velocity streaks, as is typical of the near-wall cycle. Sparse resolvent analysis gives a mode with similar characteristics, though localized in one region of the domain, with a single dominant pair of fast- and slow-moving regions, which each have a smaller spanwise extent. This indicates that sparse resolvent analysis can be applied to identify spatially-localized structures in  directions of spatial homogeneity. These sparse forcing and response modes can be interpreted as "minimal unit" structures corresponding to similar amplification as the spanwise-periodic structures identified from standard resolvent analysis. In this case, the amplification of the leading sparse modes is approximately 95\% of that of the leading standard resolvent modes. 
While not shown, here suboptimal sparse modes consist of translations of this structure along each of the critical layers.  

%%%%%%%%%%%%%%%%%%%%%%%%
% Laminar Channel Flow %
%%%%%%%%%%%%%%%%%%%%%%%%
\subsection{Space-time resolvent analysis of laminar channel flow}
\label{sec:resultsChannel}
In this section, for laminar channel (Poiseuille) flow, we first perform (standard) resolvent analysis of the space-time resolvent operator (defined in Eq.~(\ref{eq:wallNormalResolventwithT})) and show results consistent with those from standard space-only resolvent analysis (defined in Eq.~(\ref{eq:wallNormalResolvent})).%  in Sec.~\ref{sec:spaceTime_channel}.
 
We next  applying the sparsity-promoting variant on the space-time  resolvent operator to find structures that are localized in both space and time. % dimensions in Sec.~\ref{sec:sparseSpaceTime_channel}. 
Here and throughout, time is implicitly nondimensionalized by the maximum streamwise flow speed, $U_0$, and channel half-height, $h$. The numerical domain $[-h,h]\times[0,\tau)$ is discretized using 101 and 201 collocation points in the wall-normal and time dimensions, respectively.

\begin{figure}[ht!]
\vspace{-0.2cm}
\centering
\subfloat{\hspace*{-2.2cm}\includegraphics[width= 1.25\textwidth]{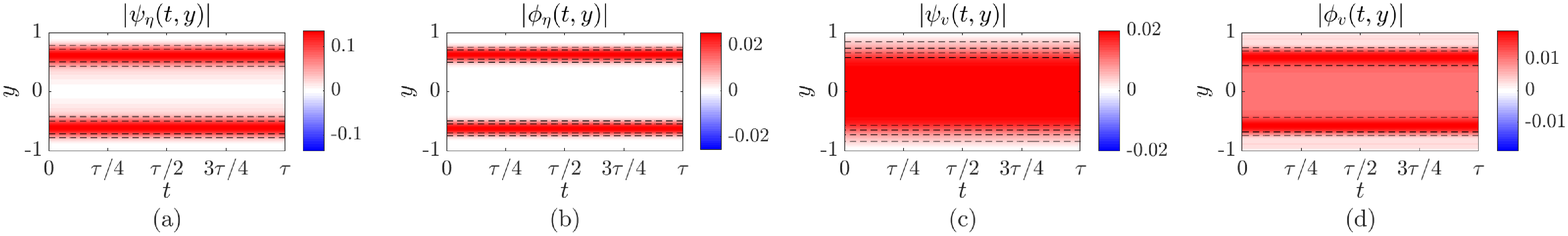}} \\
\vspace{-0.4cm}
\subfloat{\hspace*{-2.2cm}\includegraphics[width= 1.25\textwidth]{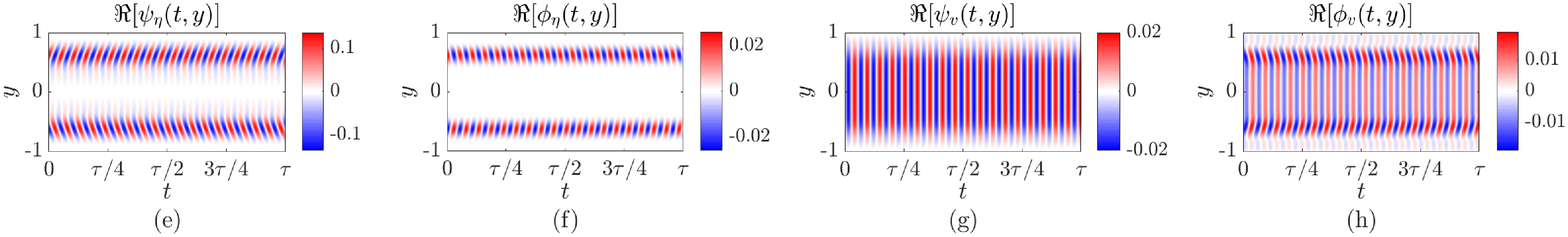}} 
\vspace{0.2cm}
\caption{Amplitude (top row) and real component (bottom row) of the leading resolvent response ($\psi$) and forcing ($\phi$) modes in wall-normal vorticity ($\eta$) and velocity ($v$) of the space-time resolvent operator for channel flow with $Re=1500$, $k_x = 2$, $k_z =1$ and a dimensionless time horizon of $\tau=100$.
}
\label{fig:spaceTime_channel}
\end{figure}

\begin{figure}[ht!]
\centering {
\vspace{-0.4cm}
{\hspace*{0cm}\includegraphics[width= 1\textwidth]{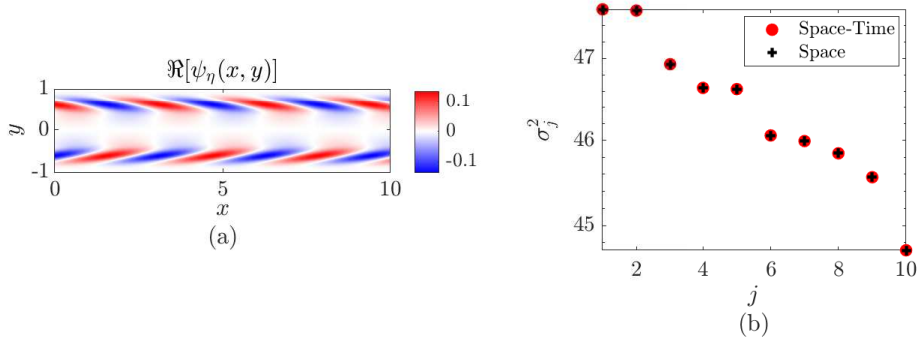}} }
\vspace{-1cm}
\caption{(a) Leading response mode in wall-normal vorticity ($\eta$) transformed to the physical domain at $t=0$; (b) first $10$ singular values $\sigma$ of the space-time resolvent operator for channel flow with the parameters indicated in Fig.~\ref{fig:spaceTime_channel} (red), and the leading singular values of the standard resolvent operator with $\omega = k {2\pi}/{\tau}$ where $k\in\{-50,...,50\}$ (black). 
}
\label{fig:spaceTime_singVals_1d2d}
\end{figure}

%\subsubsection{Space-time resolvent analysis of laminar channel flow}
\label{sec:spaceTime_channel}
Fig.~\ref{fig:spaceTime_channel} shows the wall-normal velocity ($v$) and vorticity ($\eta$) components of the leading resolvent forcing and response modes for laminar channel flow, with temporal domain %$t\in[0,\tau=100)$
$t\in[0,\tau)$ with $\tau=100$, for a representative set of parameters $k_x h=2$, $k_z h= 1$, and outer Reynolds number $Re = h U_0/\nu = 1500$. Unlike standard space-only resolvent analysis, we emphasize that here we do not specify a temporal frequency, but rather identify resolvent modes that are functions of both space and time.
Here we identify modes that have constant amplitude in time but oscillating phase, consistent with Fourier modes. The modes shown have vorticity response components localized near the critical layer (determined by the inferred temporal frequency and streamwise wavenumber). The forcing and response modes tilt in opposite directions, consistent with amplification through the Orr mechanism \cite{orr1907stability,jimenez2013linear} (note that the direction of inclination is opposite to what would be observed if the horizontal axis was $x$ rather than $t$, as is shown in Fig.~\ref{fig:spaceTime_singVals_1d2d}(a)). The wall-normal velocity response mode components consist of upright structures extending across the full height of the domain. Regions with $v$ directed away from the wall corresponds to regions of low $\eta$ and vice-versa, consistent with the lift-up mechanism transporting momentum in the $y$-direction. 

These modes should be identical to those identified using standard resolvent analysis across all permissible frequencies (\textit{i.e.} $\omega = k 2\pi/\tau$ with $k\in \mathbb{Z}$%, where $\tau$ is the length of the time domain
). To confirm this, in Fig.~\ref{fig:spaceTime_singVals_1d2d}(b) we compare the leading singular values of the temporally-localized resolvent operator with the maximal singular values obtained from traditional resolvent analysis over a range of permissible frequencies. The agreement between the singular values supports the premise that, for a stationary base/mean flow, space-time resolvent analysis indeed compiles the results of performing standard resolvent analysis across a range of permissible frequencies.

\begin{figure}[ht!]
\vspace{-0.2cm}
\centering
\vspace{0cm}
\centering
\subfloat{\hspace*{-2.2cm}\includegraphics[width= 1.25\textwidth]{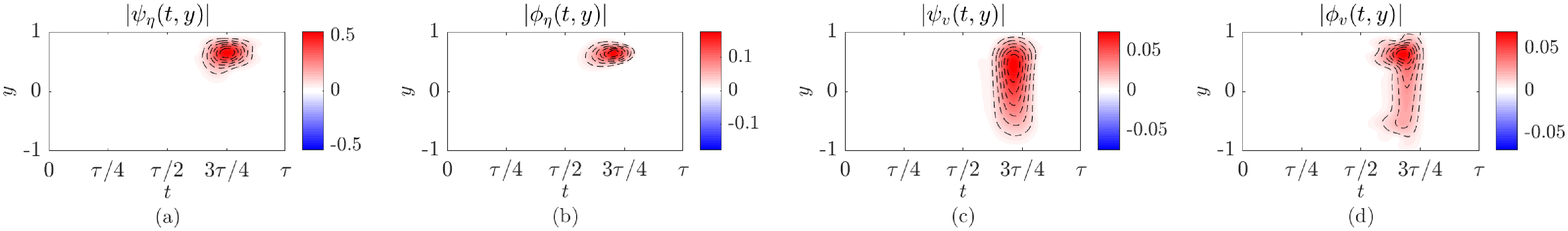}} \\
\vspace{-0.5cm}
\subfloat{\hspace*{-2.2cm}\includegraphics[width= 1.25\textwidth]{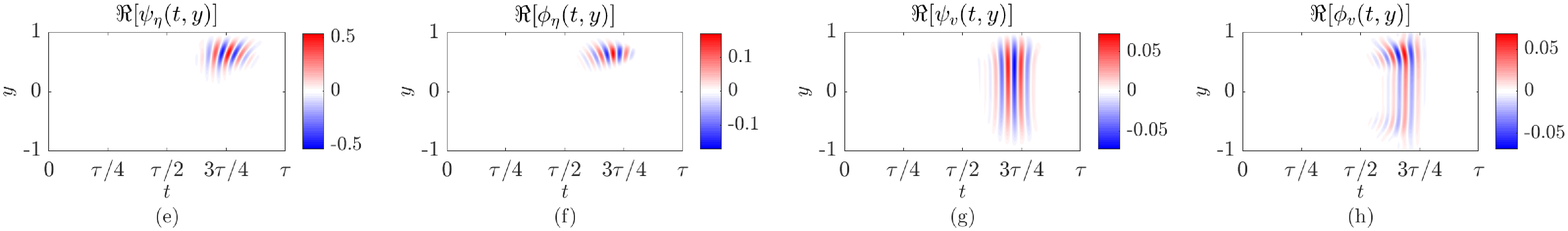}} 
\vspace{0cm}
\caption{Amplitude (top row) and real component (bottom row) of the leading regularized sparse resolvent response ($\psi$) and forcing ($\phi$) modes in wall-normal vorticity ($\eta$) and velocity ($v$) of the spatio-temporal resolvent operator for channel flow with the parameters indicated in Fig.~\ref{fig:spaceTime_channel} and a sparsity parameter of $\gamma=0.01$.
}
\label{fig:spaceTime_channel_sparse}
\end{figure}

\begin{figure}[ht!]
\centering {
\vspace{-0.3cm}
{\hspace*{-2.2cm}\includegraphics[width= 1.20\textwidth]{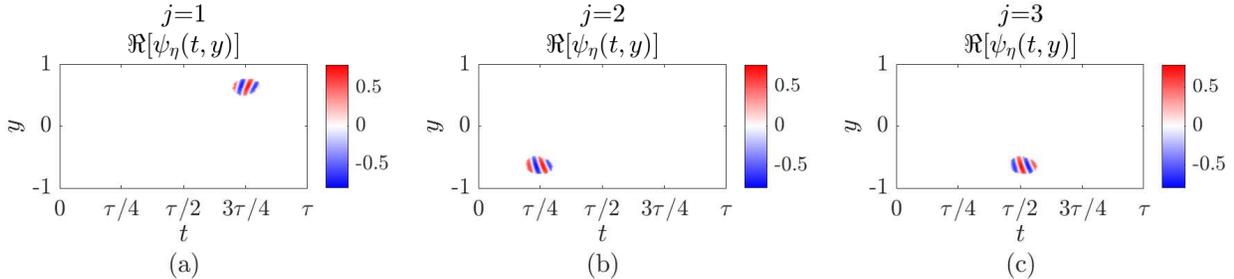}} }
\vspace{-0.4cm}
\caption{Real component of the first $j=\{1,2,3\}$ non-regularized (raw) sparse response modes in wall-normal vorticity ($\eta$) of the spatio-temporal resolvent operator for channel flow with the parameters indicated in Fig.~\ref{fig:spaceTime_channel} and a sparsity parameter of $\gamma=0.01$.
}
\label{fig:spaceTime_channel_sparse_raw}
\end{figure}

%%%%%
%\subsubsection{Sparse space-time resolvent analysis on Laminar Channel Flow}
%\label{sec:sparseSpaceTime_channel}
We now consider sparse space-time resolvent analysis for the same configuration and parameters. Fig.~\ref{fig:spaceTime_channel_sparse} depicts the sparse leading resolvent mode that were obtained for $\gamma=0.01$, allowing direct comparison with the non-sparse resolvent modes presented in Fig.~\ref{fig:spaceTime_channel}. %The first four modes here are just a temporal and spatial translations of the first. 
These response modes look similar to those shown for standard (non-sparse) space-time resolvent analysis, except for being localized both in time and space. 
 The vorticity components of the forcing and response modes are again concentrated in a localized region near the critical layer corresponding to the frequency of oscillation of the phase of the modes, while the wall-normal velocity components extend over a wider region of the domain. 
The forcing modes show a slightly greater degree of spatial localization than the response modes, particularly for the vorticity component. Note that the forcing and response modes shown in Fig.~\ref{fig:spaceTime_channel_sparse} are computed using Eqs.~(\ref{eq:forcing})-(\ref{eq:responseupdated}), rather than being the direct output of the sparse optimization method. For comparison, we additionally show the raw output of optimizing Eq.~(\ref{eq:minsvdsparse}) in Fig.~\ref{fig:spaceTime_channel_sparse}. Only the vorticity component is shown, since for this choice of sparsity parameter $\gamma$, the wall-normal velocity component is zero. Comparison between Figs.~\ref{fig:spaceTime_channel_sparse} and \ref{fig:spaceTime_channel_sparse_raw} shows how applying Eqs.~(\ref{eq:forcing})-(\ref{eq:responseupdated}) modifies the raw sparse response modes to give structures more closely resembling the leading standard resolvent modes, at the expense of sparsity. In particular, in this case applying Eqs.~(\ref{eq:forcing})-(\ref{eq:responseupdated}) gives a response mode that spans over a less localized region, as well as nonzero wall-normal velocity components. Note that for both standard and sparse resolvent analysis, the wall-normal vorticity has a much larger response for these parameters.
%in both the upper and lower critical layers, even though the raw sparse modes shown in Fig.~\ref{fig:spaceTime_channel_sparse_raw} are localized at only one of the critical layer locations.

\begin{figure}[ht!]
\centering {
\vspace{0cm}
{\hspace*{-2.2cm}\includegraphics[width= 1.25\textwidth]{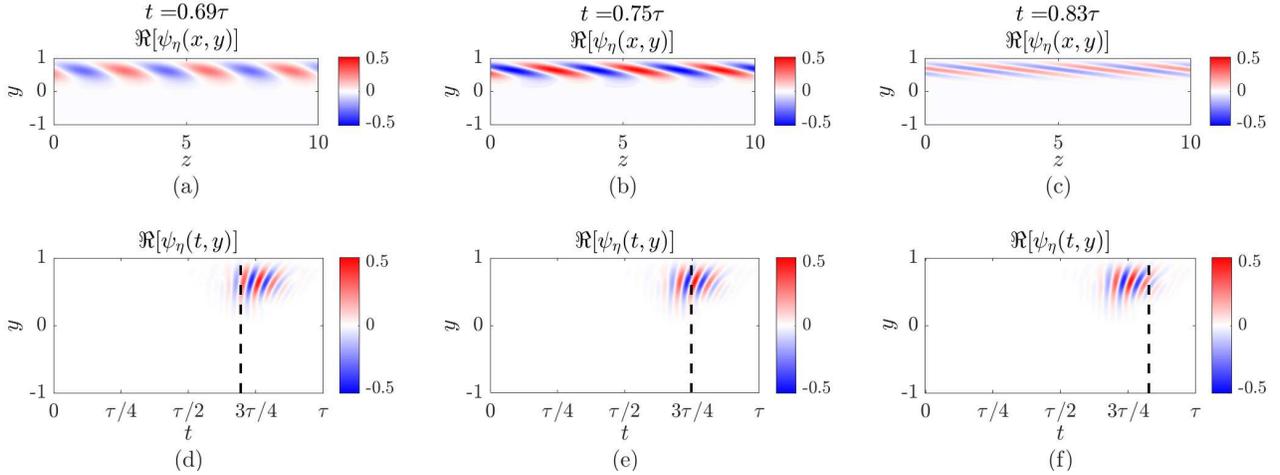}} }
\vspace{-0.4cm}
\caption{Real part of the wall-normal vorticity ($\eta$) component of the instantaneous leading response mode in the physical domain at $t=\{0.69\tau, 0.75\tau, 0.83\tau\}$ (top row) and as a function of space $y$ and time $t$ with $t \in [0,\tau)$ (bottom row) of the spatiotemporal resolvent operator for channel flow with the parameters indicated in Fig.~\ref{fig:spaceTime_channel} and a sparsity parameter of $\gamma=0.01$. The corresponding time instances are marked with a black dashed line in the lower subplots.
}
\label{fig:physical_sparse_channel}
\end{figure}

\begin{figure}[ht!]
\centering {
\vspace{-0.4cm}
{\hspace*{-1.35cm}\includegraphics[width= 1.15\textwidth]{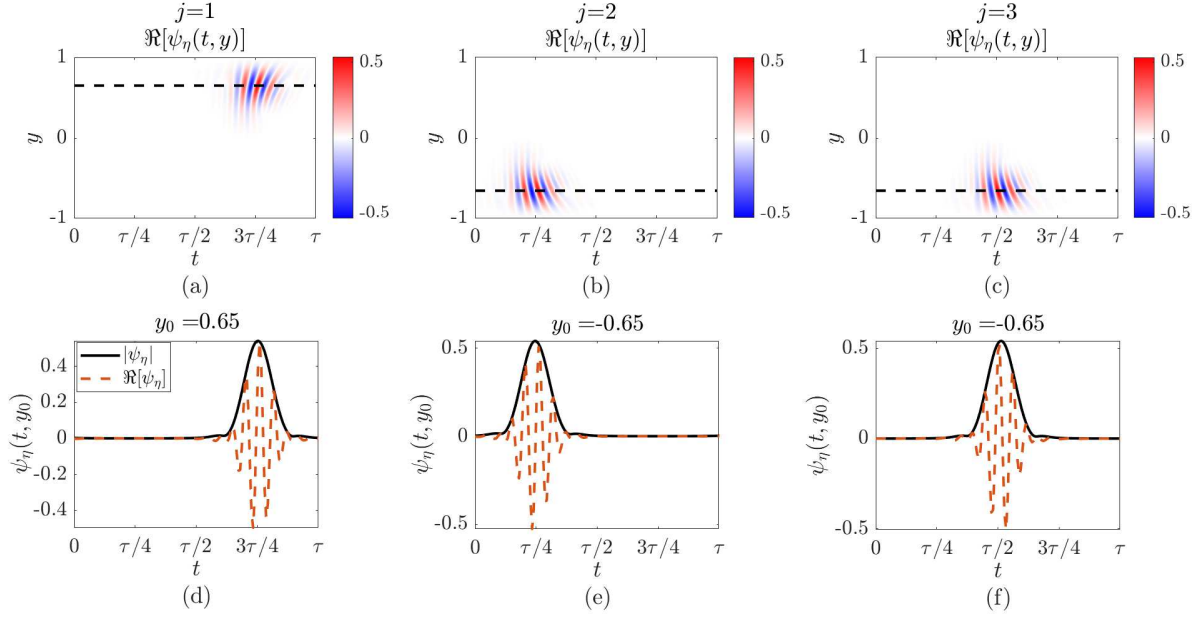}} }
\vspace{-.6cm}
\caption{(a)-(c) Real part of the wall-normal vorticity ($\eta$) component of the first three $j=\{1,2,3\}$ leading sparse response modes of the spatio-temporal resolvent operator for channel flow with the parameters indicated in Fig.~\ref{fig:spaceTime_channel}; (d)-(f) cross-sections along the $t$-axis of modes (a)-(b) at the spatial locations of maximum mode amplitude, $y_0$. 
}
\label{fig:crossSectionsT_channel}
\end{figure}

The behavior of the time-localized modes is further studied by considering the strucutre of the modes at certain instances of time. 
%of the observed sparse structures becomes even more evident when examining their physical counterpart at several time instances.
In Fig.~\ref{fig:physical_sparse_channel}, we visualize the leading sparse response mode in vorticity along the streamwise axis at three different  locations in time. %These particular instances are representative of the sparsity of these coherent structures,
These show how the mode amplitude grows and then decays over time, while the streamwise inclination of the modes increases.
The time-localization of these structures can be observed more directly in Fig.~\ref{fig:crossSectionsT_channel}, which depicts the cross-sections along the $t$-axis of the first three sparse  response modes in vorticity at the spatial locations of largest mode amplitude. These cross sections show that the identified localized temporal functions appear to resemble Gaussian envelopes, with approximately constant phase gradient. 
 This could provide evidence for a natural wavelet template for performing time-localized resolvent analysis with a prescribed wavelet basis. 
%This result could certainly support the possibility introducing wavelet-based methods to study these particular coherent structures,
The structure of these temporal modes also suggests a connection with similar wavepacket templates that have been shown to closely approximate spatial resolvent modes~\cite{dawson2019shape,dawson2020ijhff}.
%as well as the use of analytical approximations to fit the parameters of a suitably-defined template (in connection to the work described in Ref.~\cite{dawson2019shape}).

% Mention time shift between forcing and response

%%%%%%%%%%%%%%%%%%%%%%%
% Turbulent Stokes BL %
%%%%%%%%%%%%%%%%%%%%%%%
\subsection{Space-time resolvent analysis of a turbulent Stokes boundary layer}
\label{sec:resultsStokesBL}
%, in Sec.~\ref{sec:sparseSpaceTime_StokesBL}.
% The DNS data pertaining the mean velocity profile and its derivatives was obtained with the parameters indicated in Fig.~\ref{fig:StokesBL}. Note how for this particular system, the application of the space-time formulation of resolvent analysis could be particularly useful since here the use of standard resolvent analysis (involving a Fourier transform in time) would not be reasonable due to the unsteadiness of the mean. 

\begin{figure}[ht!]
\vspace{-0.5cm}
\centering
\vspace{0cm}
\centering
\subfloat{\hspace*{0cm}\includegraphics[width= 0.5\textwidth]{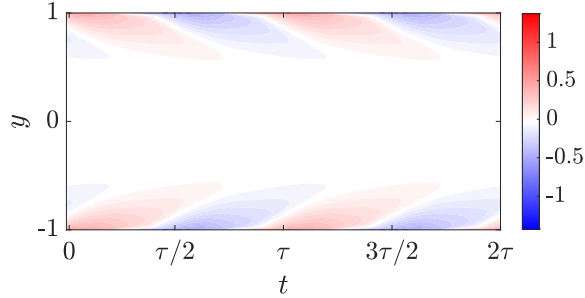}} \\ 
\vspace{-0.3cm}
\caption{Turbulent mean streamwise velocity profile of a turbulent Stokes boundary layer with $Re_{\Omega}=1500$, for two periods of oscillation.
%$\omega=0.08$, $\nu=2.5x10^{-5}$, $U_0=1.5$ and $d_t=3.927x10^{-4}$ for a full oscillating period of $\tau=2\pi/\omega=7.854$.
}
\label{fig:StokesBL}
\end{figure}

%\subsubsection{Space-time resolvent analysis of a turbulent Stokes Boundary Layer}
%\label{sec:spaceTime_StokesBL}
We now consider a system where the mean velocity profile varies in time. In particular, we consider a Stokes boundary layer configuration, where we again have flow between two parallel plates, but without an imposed pressure gradient, and where the boundaries move with a velocity 
\begin{equation}
    U_w(t) = U_{{0}}\cos(\Omega t)
\end{equation}
We consider a Reynolds number based on the Stokes boundary layer thickness $\delta_\Omega = \sqrt{2\nu/\Omega}$ of $Re_\Omega = U_{0} \delta_\Omega/\nu = 1500$. At this Reynolds number, the flow is intermittently turbulent \cite{akhavan1991investigation,verzicco1996direct,vittori1998direct,costamagna2003coherent}. The time-periodic mean velocity profile  for this configuration is shown in Fig.~\ref{fig:StokesBL}.

In this section, we again first consider space-time resolvent analysis with the standard optimization problem,  before considering the sparsity-promoting variant  on the observed time-resolved structures, allowing for the identification of time-localized structures corresponding to large linear amplification. Here and throughout, time is implicitly nondimensionalized by the characteristic velocity, $U_0$, and channel half-height, $h$. The numerical domain $[-h,h]\times[0,\tau)$ is discretized using 121 and 201 collocation points in the wall-normal and time dimensions, respectively.

\begin{figure}[ht!]
\vspace{0cm}
\centering
\vspace{0cm}
\centering
\subfloat{\hspace*{-2.2cm}\includegraphics[width= 1.25\textwidth]{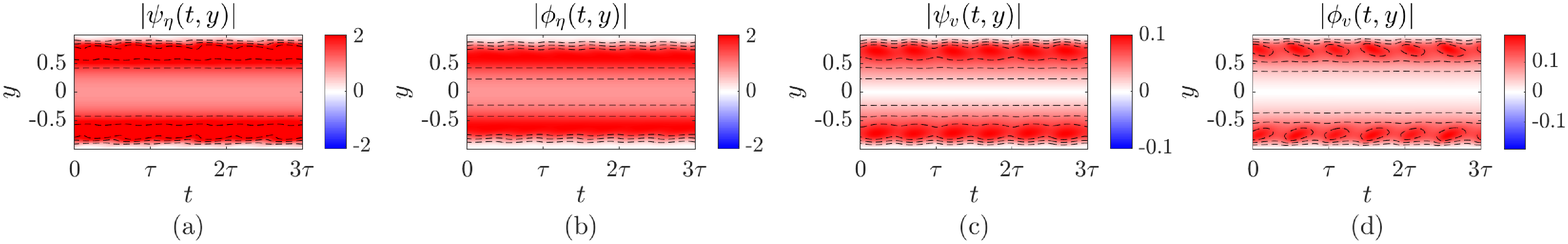}} \\
\vspace{-0.6cm}
\subfloat{\hspace*{-2.2cm}\includegraphics[width= 1.25\textwidth]{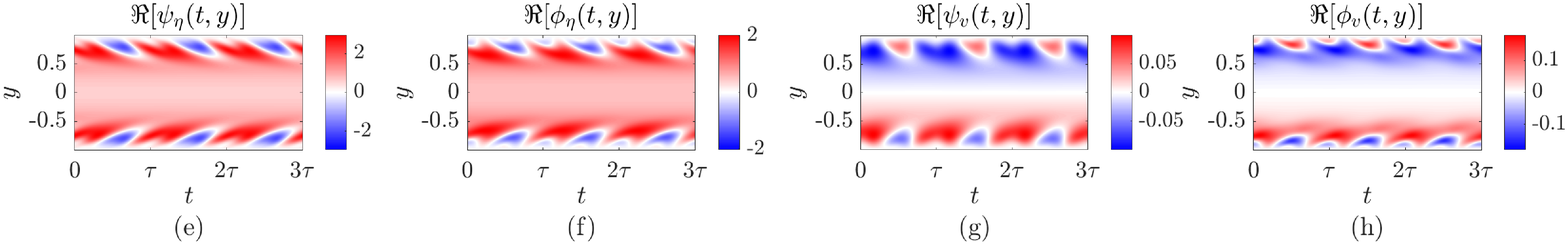}} 
\vspace{0cm}
\caption{Amplitude (top row) and real component (bottom row) of the leading resolvent response ($\psi$) and forcing ($\phi$) modes in wall-normal vorticity ($\eta$) and velocity ($v$) of the spatio-temporal resolvent operator for a turbulent Stokes boundary layer with $Re_{\Omega}=1500$, $k_x h= 3\pi$, and $k_z h =3\pi$, %$\omega=0.08$ and a time horizon $T=3\tau$, corresponding to $n=3$
for three periods of mean flow oscillation. 
}
\label{fig:spaceTime_StokesBL}
\end{figure}

Fig.~\ref{fig:spaceTime_StokesBL} shows the wall-normal velocity and vorticity components of the leading resolvent forcing and response modes for a turbulent Stokes boundary layer, over a time domain that spans over three boundary layer cycles. Throughout this section we consider streamwise and spanwise wavenumbers $k_x h = k_z h= 3\pi$.
Although this system is time-periodic, the fact that it is not statistically-stationary means that each space-time resolvent mode does not necessarily correspond to a single Fourier mode in time. Indeed, the amplitude of the leading modes depicted in  Fig.~\ref{fig:spaceTime_StokesBL}(a)-(d) is not exactly constant in time. Comparison between the real components of the velocity and vorticity of the response modes (Fig.~\ref{fig:spaceTime_StokesBL}(e)-(h))
indicates that wall-normal velocity directed  away from (towards) the wall corresponds to positive (negative) values of wall-normal vorticity, in turn indicative of positive (negative) streamwise velocity fluctuations, suggesting the presence of the lift-up mechanism transporting momentum away from the side-walls.

\begin{figure}[ht!]
\vspace{0cm}
\centering
\vspace{0cm}
\centering
\subfloat{\hspace*{-2.2cm}\includegraphics[width= 1.25\textwidth]{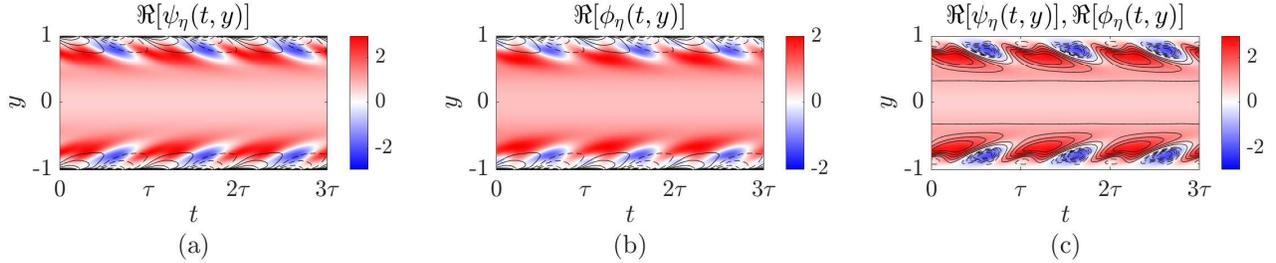}} \\ 
\vspace{0cm}
\caption{Real component of the leading response ($\psi$) (a) and forcing ($\phi$) (b) modes in vorticity ($\eta$) and contour levels of the streamwise turbulent mean velocity profile $U_0$; (c) real component of the leading response mode in vorticity and contour levels of the real component leading forcing mode in vorticity. The parameters considered here are the same as indicated in Fig.~\ref{fig:spaceTime_StokesBL}. Here  solid and dashed lines indicate positive and negative contour levels, respectively.
}
\label{fig:forcingResponse_mean_StokesBL}
\end{figure}

%suggests that the lift-up mechanism manifests in this system as well, transporting momentum away from the side-walls.
The real components of this forcing and response mode oscillate with a period equal to the mean.
 To study the relationship between the mean and leading resolvent modes more directly, we show in Fig.~\ref{fig:forcingResponse_mean_StokesBL}(a)-(b) a comparison between the $\eta$-component of the leading resolvent forcing and response mode with the mean velocity profile.  We observe a phase shift of approximately a quarter of a period between the mean and resolvent mode contours, for both the forcing and response. In contrast, for these parameters there is little phase shift between the forcing and response mode components, as shown in Fig.~\ref{fig:forcingResponse_mean_StokesBL}(c).

%More specifically, we show that the shift between both forcing and response modes in vorticity and the turbulent mean are shifted approximately less than a quarter of a period in Fig.~\ref{fig:forcingResponse_mean_StokesBL} (a)-(b). We attribute this to viscous effects, due to the rapid changes of the mean and sharp gradients near the side-walls. In terms of the synchronicity between forcing and response, it can be seen in Fig.~\ref{fig:forcingResponse_mean_StokesBL} (c) that there is little to none time-shift between both coherent structures. A phase-shift is indeed present, however it is significantly smaller compared to the channel case, which suggests that the contribution of the Orr mechanism may not be as significant here.

\begin{figure}[ht!]
\centering {
\vspace{0cm}
{\hspace*{-2.2cm}\includegraphics[width= 1.25\textwidth]{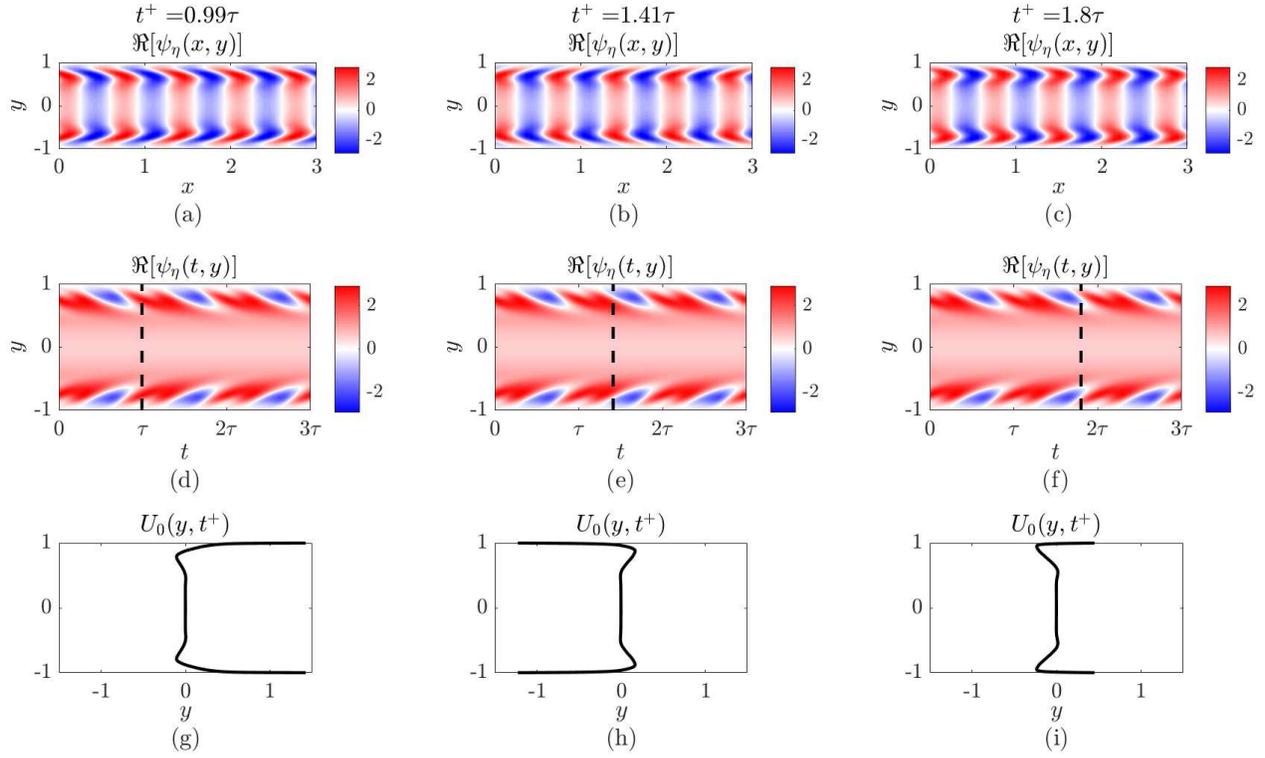}} }
\vspace{-0.5cm}
\caption{(a)-(c) Real part of the wall-normal vorticity ($\eta$) component of the instantaneous leading response mode ($\psi$) in the physical domain at $t=\{0.99\tau, 1.41\tau, 1.80\tau\}$; (d)-(f) the same mode components plotted in space ($y$) and time ($t$), with the parameters indicated in Fig.~\ref{fig:spaceTime_StokesBL}. The time instances shown in (a)-(c) are marked with a black dashed in (d)-(f). Corresponding instantaneous turbulent mean profiles are shown in (g)-(i).
}
\label{fig:physical_BL}
\end{figure}

Lastly, in order to %contextualize
further investigate
the time-evolving structures identified by this analysis, we present the corresponding physical mode as a function of $x$ and $y$ (by undoing the Fourier transform in the streamwise direction) of the leading space-time response mode in vorticity along with the instantaneous turbulent mean at three time instances in Fig.~\ref{fig:physical_BL}. The observed periodicity of the space-time modes manifests as the changing phase of the physical modes in the vicinity of the side-walls, following the temporal evolution of the turbulent mean. Interestingly, the inclination of the modes shown in Fig.~\ref{fig:physical_BL}(a)-(c) is in the opposite direction to that which would be expected for a stationary mean profile, where the response modes tilt in the same direction as the mean profile (e.g. as consistent with the Orr mechanism). %This phase shift will be the study of further investigation.

%Additionally, there is a resemblance between the contour levels of the spatial modes and the instantaneous mean, which could be indicative of the high dissipation levels produced by the pronounced velocity gradients in the near-wall regions. 

\begin{figure}[ht!]
\vspace{0cm}
\centering
\vspace{0cm}
\centering
\subfloat{\hspace*{-2.2cm}\includegraphics[width= 1.25\textwidth]{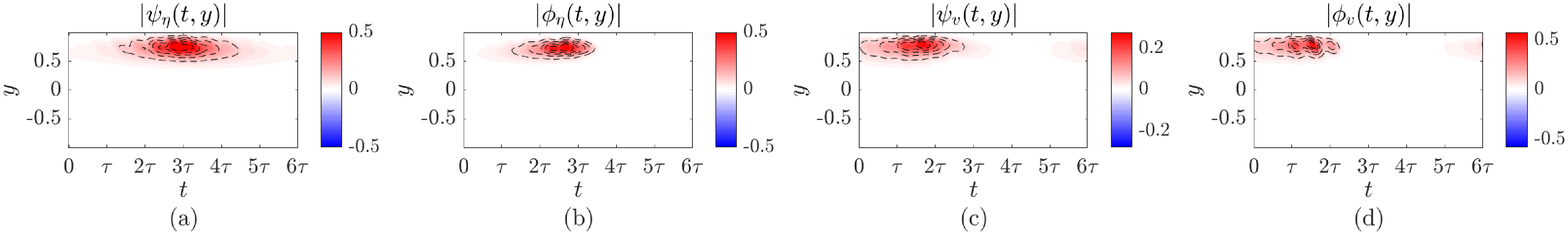}} \\
\vspace{-0.4cm}
\subfloat{\hspace*{-2.2cm}\includegraphics[width= 1.25\textwidth]{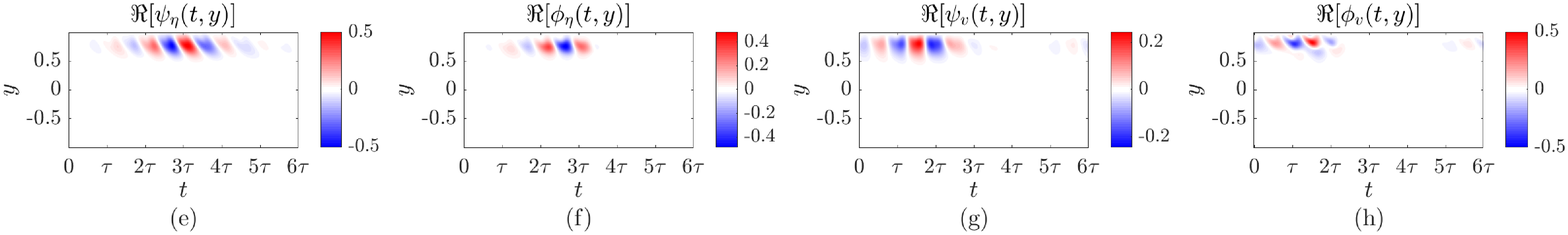}} 
\vspace{0cm}
\caption{Amplitude (top row) and real component (bottom row) of the leading regularized sparse space-time resolvent response ($\psi$) and forcing ($\phi$) modes in wall-normal vorticity ($\eta$) and velocity ($v$) %of the spatio-temporal resolvent operator
for a turbulent Stokes boundary layer with the parameters indicated in Fig.~\ref{fig:spaceTime_StokesBL}, over a time domain consisting of 6 periods of mean flow oscillation. %$T=6\tau=47.124$, corresponding to $n=6$ oscillating cycles, and a sparsity parameter of $\gamma=0.01$.
}
\label{fig:spaceTime_StokesBL_sparse}
\end{figure}

%\subsubsection{Sparse space-time resolvent analysis on turbulent Stokes Boundary Layer}
%\label{sec:sparseSpaceTime_StokesBL}
We now consider sparse space-time resolvent analysis of the same configuration. 
Fig.~\ref{fig:spaceTime_StokesBL_sparse} shows the leading sparse forcing and response resolvent mode components  obtained for a sparsity parameter $\gamma=0.01$ (in analogy to Fig.~\ref{fig:spaceTime_StokesBL} for the non-sparse case). In order to best highlight the sparsity of the observed structures, in this case the time domain spans six periods of the mean flow. 
The sparse resolvent method identifies oscillating structures that are localized both in space and time, though for these parameters the mode components each span several oscillation periods. We observe as expected that the forcing mode components tend to precede the corresponding response mode components. In addition, the wall-normal velocity  components precede the wall-normal vorticity, again indicating an energy transfer pathway consistent with the lift-up mechanism.

%Here, while the sparse PCA identifies oscillating structures that are localized both in space and time, a priori they do not seem to represent a localized version of the non-sparse leading mode depicted in Fig.~\ref{fig:spaceTime_StokesBL}. However, both sparse and non-sparse structures in fact share some common features, such as the oscillatory behavior and the phase difference between forcing and response modes, perhaps attributed to the periodicity of the mean.  

\begin{figure}[ht!]
\vspace{-0.3cm}
\centering
\vspace{0cm}
\centering
\subfloat{\hspace*{-2.2cm}\includegraphics[width= 1.25\textwidth]{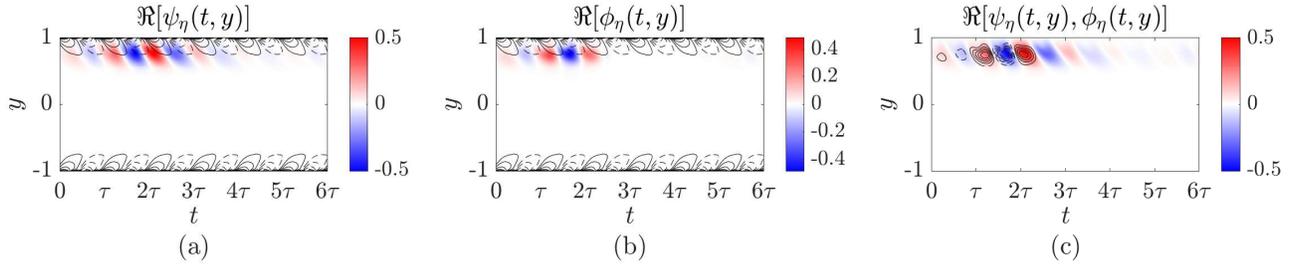}} \\ 
\vspace{0.3cm}
\caption{Real component of the leading response (a) and forcing (b) modes in vorticity and contour levels of the turbulent mean velocity profile; (c) real component of the leading response mode in vorticity and contour levels of the real component leading forcing mode in vorticity. The parameters considered here are the same as indicated in Fig.~\ref{fig:spaceTime_StokesBL_sparse}. Here the positive and negative contour levels are reprented as solid and dashed lines, respectively. 
}
\label{fig:forcingResponse_mean_StokesBL_sparse}
\end{figure}

Fig.~\ref{fig:forcingResponse_mean_StokesBL_sparse}(a-b) shows the relationship between the $\eta$-component of the forcing and response modes with the mean velocity. Again (\textit{c.f.} Fig.~\ref{fig:forcingResponse_mean_StokesBL}) we observe a quarter-period shift between the resolvent modes and mean. Fig.~\ref{fig:forcingResponse_mean_StokesBL_sparse}(c) confirms that the $\eta$-component of the forcing and response modes are again in-phase for these parameters.  These findings suggest that the sparsity-promoting method is identifying a time-localized version of the same mechanism that was identified using the standard optimization method.

%Furthermore, it is possible that the contribution of the Orr mechanism may be slightly higher when these structures are localized, as the phase difference between forcing and response is somewhat higher. According to Fig.~\ref{fig:forcingResponse_mean_StokesBL_sparse} (c), while there does not seem to be a time shift between forcing and response, the phase difference is noticeable. \textbf{Thus, the periodicity of the turbulent mean could be the reason behind the suppression of the Orr mechanism}. % could this be why? or am i making it up?  I'm not sure if the orr mechanism is completely suppressed, but rather that there is a phase shift effect that needs to be accounted for. 
%Additionally, time shift observed in Fig.~\ref{fig:forcingResponse_mean_StokesBL_sparse}(a)-(b) is consistent with the results presented for the non-sparse system - which are likely attributed to sharp velocity gradients and the consequent generation of shear stresses.  % THIS IS INTERESTING, WE SHOULD DISCUSS FURTHER. 

%The forcing structures show a greater degree of localization in both components, in agreement to the result observed in Sec.~\ref{sec:spaceTime_channel}, and due to the sequence of steps within the algorithm. It is interesting to note, however, that while the regularized sparse modes span over multiple cycles, the unregularized (or raw) modes identified by the sparse PCA merely span for about an oscillating period (see Fig.~\ref{fig:spaceTime_StokesBL_sparseRaw}). It seems as though the method is able to recognize the periodicity of the turbulent mean (as well as the resulting coherent structures), and it encapsulates that feature on the sparse results. 

% \begin{figure}[ht!]
% \centering {
% \vspace{0.5cm}
% {\hspace*{-2.2cm}\includegraphics[width= 1.25\textwidth]{Figures/resolventModes_Dt_BL_AbsReal_Re1500_kx3pi_kz3pi_6cycles_SPARSE_raw.eps}} }
% \vspace{-0.4cm}
% \caption{Real component of the first $j=\{1,2,3\}$ non-regularized (raw) sparse response modes in wall-normal vorticity ($\eta$) of the spatio-temporal resolvent operator for a turbulent Stokes boundary layer with the parameters indicated in Fig.~\ref{fig:spaceTime_StokesBL_sparse}.
% }
% \label{fig:spaceTime_StokesBL_sparseRaw}
% \end{figure}

\begin{figure}[ht!]
\centering {
\vspace{0.2cm}
{\hspace*{-2.2cm}\includegraphics[width= 1.25\textwidth]{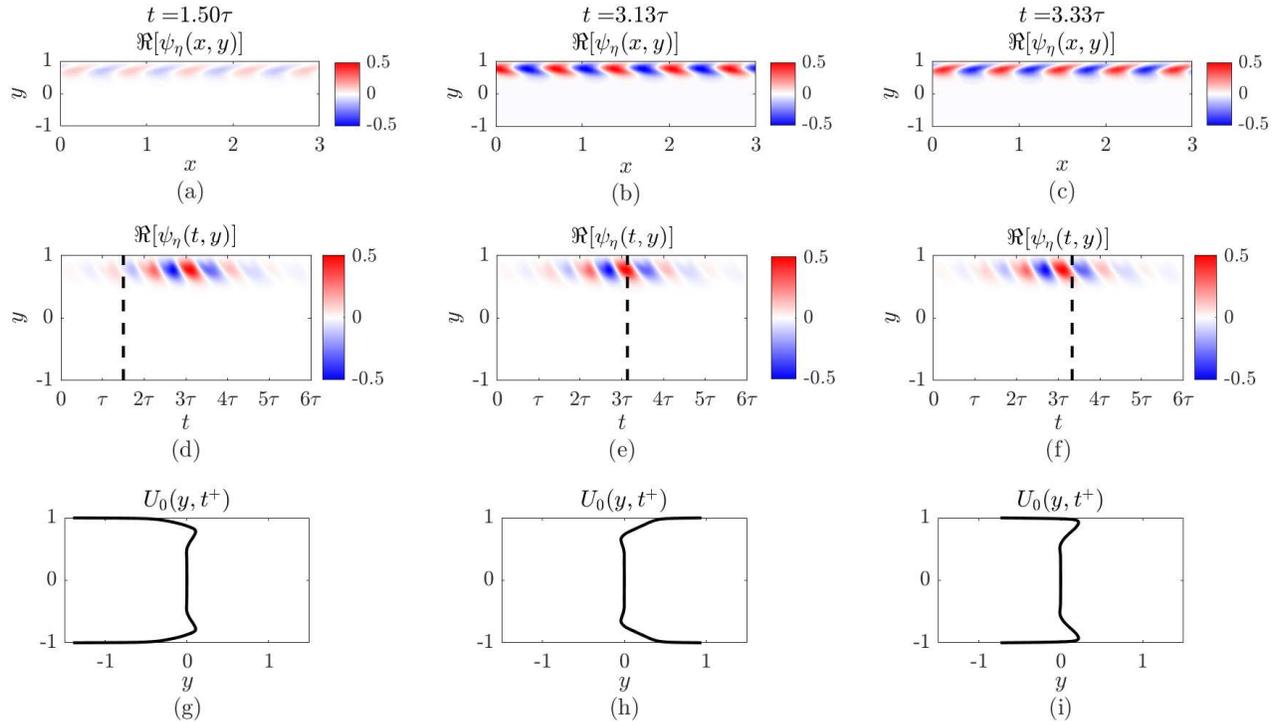}} }
\vspace{-0.5cm}
\caption{Real part of the wall-normal vorticity ($\eta$) component of the instantaneous leading response mode in the physical domain at $t=\{1.50\tau, 3.13\tau, 3.33\tau\}$ (top row) and as a function of space $y$ and time $t$ of the spatiotemporal resolvent operator for a turbulent Stokes boundary layer with the parameters indicated in Fig.~\ref{fig:spaceTime_StokesBL_sparse}. The time instances of interested are marked with a black dashed line over the space-time modes. Instantaneous turbulent profiles (g)-(i).}
\label{fig:physical_sparse_Stokes}
\end{figure}

The temporal evolution of the sparse response mode shown in Fig.~\ref{fig:physical_sparse_Stokes} also shows similar behavior to that observed for the non-sparse modes, with a mode inclination angle in the $x-y$ plane that appears to tilt in the opposite direction to the mean velocity profile at a given instant in time.

%: phase-dependency of the modes with respect to the temporal evolution of the mean velocity profile, as well as high amplification near the side-walls due to high shear stresses.  

\begin{figure}[ht!]
\centering {
\vspace{-0.4cm}
{\hspace*{-1.35cm}\includegraphics[width= 1.15\textwidth]{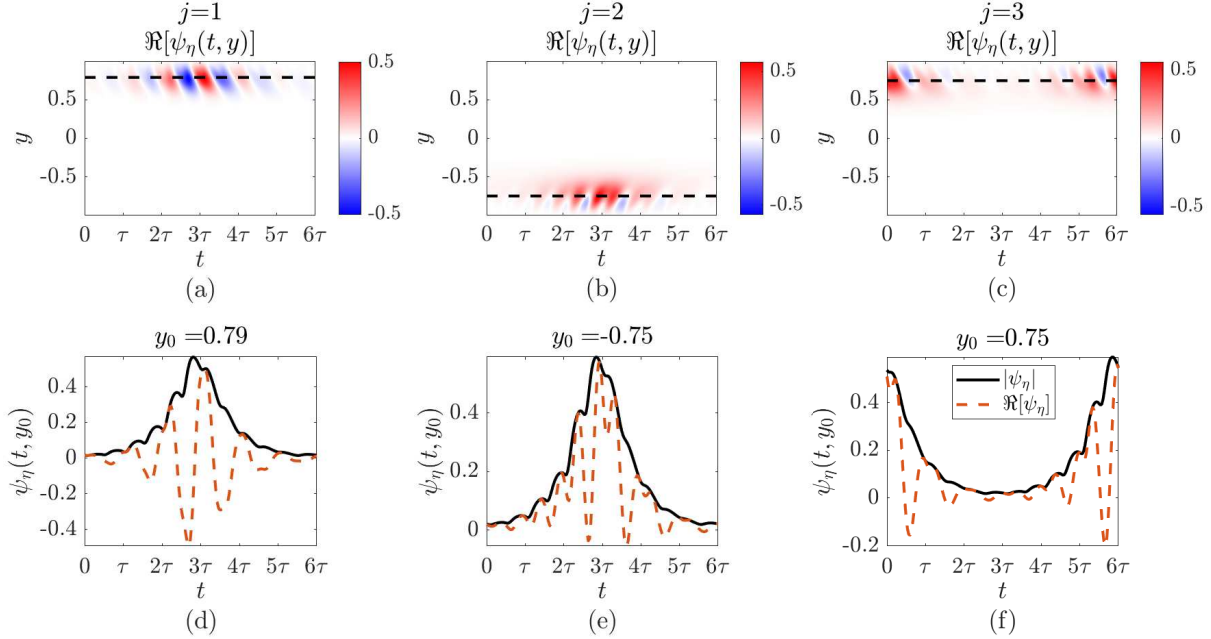}} }
\vspace{-.6cm}
\caption{(a)-(c) Real part of the wall-normal vorticity ($\eta$) component of the first three $j=\{1,2,3\}$ leading sparse response modes of the spatiotemporal resolvent operator for a turbulent Stokes boundary layer with the parameters indicated in Fig.~\ref{fig:spaceTime_StokesBL_sparse}; (d)-(f) cross-sections along the $t$-axis of modes (a)-(b) at the locations of corresponding maximum amplitude, $y_0$. 
}
\label{fig:crossSectionsT_StokesBL_sparse}
\end{figure}

Lastly, the temporal evolution of the sparse response modes at the locations of maximum amplitude $y_0$ are shown in Fig.~\ref{fig:crossSectionsT_StokesBL_sparse}. In contrast to the statistically-stationary mean case considered in Fig.~\ref{fig:crossSectionsT_channel}, here we observe more complex temporal envelopes, which cannot necessarily be closely approximated by a single simple prescribed template function. %This suggests that 

%suggests that one Gaussian envelope would not suffice as a compact representation of these sparse structures. Instead a superposition of number localized curves such as wavelets would constitute a better representation of these isolated effects. 

\section{Discussion and Conclusions}
In this work we have described a space-time extension of resolvent analysis, and have developed a variant of the corresponding optimization problem that can identify forcing and response modes that are sparse, either in space or both space and time. The method has been applied to several channel flow configurations, showing that it can identify modes that are sparse in (i) the wall-normal direction, (ii) the wall-normal and spanwise directions, and (iii) the wall-normal direction and in time. 
This sparsity-promoting variant incorporates %utilizes
an $l_1$-penalization term on the resolvent response modes, giving an optimization problem that can be solved using an inverse-power method applied to a corresponding nonlinear eigenproblem. 
When applied to the standard space-only resolvent operator (assuming a Fourier transform in time), the sparsity-promoting  variant identified localized modes with similar structure to their non-sparse equivalents. 

For statistically-stationary systems, we verified that the space-time extension of resolvent analysis recovered a Fourier decomposition in time. When using the sparse version, time-localized structures were isolated, while again containing many of the same features as the equivalent standard resolvent modes. These space-time forcing and response structures can be viewed as an intermediary between transient growth analysis (which considers energy amplification between two instances in time), and standard resolvent analysis.

The fact that the space-time resolvent formulation uses an operator discretized in both space and time makes it amenable for the analysis of time-evolving systems. This was explored by considering a turbulent Stokes boundary layer.  This analysis (which focused on relatively small spatial wavelengths) identified forcing and response structures that oscillate with the same frequency as the mean flow. The sparse resolvent analysis identified modes that were localized in time, but still extended beyond one period of the boundary layer. For both standard and sparse resolvent analysis, the wall-normal vorticity components of forcing and response were a quarter a period out of phase from the mean. The streamwise inclination of the leading response modes were also observed to be in the opposite direction to the mean flow at a given instance in time, in contrast to typical behavior found in statistically-stationary systems.
 Further research will include a more comprehensive study of the linear amplification properties of the turbulent Stokes boundary layer over a range of spatial scales, as well as application of the space-time resolvent methodology (both with and without sparsity promotion) to non-periodic time-varying systems.

\section*{Acknowledgments}
This work was supported by the Air Force Office of Scientific Research grant FA9550-22-1-0109. STMD and BLD thank K. Rosenberg for sharing his two-dimensional resolvent code in velocity-vorticity form.
\bibliography{Master}

\begin{thebibliography}{53}
\newcommand{\enquote}[1]{``#1''}
\providecommand{\natexlab}[1]{#1}
\providecommand{\url}[1]{\texttt{#1}}
\providecommand{\urlprefix}{URL }
\expandafter\ifx\csname urlstyle\endcsname\relax
  \providecommand{\doi}[1]{\discretionary{}{}{}https://doi.org/#1}\else
  \providecommand{\doi}[1]{\discretionary{}{}{}\urlstyle{rm}\url{https://doi.org/#1}}\fi

\bibitem[{Schmid and Henningson(2001)}]{Schmid_Henningson}
Schmid, P., and Henningson, D.~S., \emph{Stability and Transition in Shear
  Flows}, Springer, New York, NY, 2001.

\bibitem[{McKeon and Sharma(2010)}]{mckeon2010resolvent}
McKeon, B.~J., and Sharma, A.~S., \enquote{A critical-layer framework for
  turbulent pipe flow,} \emph{Journal of Fluid Mechanics}, Vol. 658, 2010, pp.
  336--382.

\bibitem[{Holmes et~al.(2012)Holmes, Lumley, Berkooz, and
  Rowley}]{holmes2012pod}
Holmes, P., Lumley, J.~L., Berkooz, G., and Rowley, C.~W., \emph{Turbulence,
  coherent structures, dynamical systems and symmetry}, Cambridge University
  Press, 2012.

\bibitem[{Lumley(1967)}]{lumley1967}
Lumley, J.~L., \enquote{The structure of inhomogeneous turbulent flows,}
  \emph{Proceedings of the International Colloquium on the Fine Scale Structure
  of the Atmosphere and its Influence on Radio Wave Propagation}, edited by
  A.~M. Yaglam and V.~I. Tatarsky, Doklady Akademii Nauk SSSR, Moscow, Nauka,
  1967.

\bibitem[{Towne et~al.(2018)Towne, Schmidt, and Colonius}]{towne2018spectral}
Towne, A., Schmidt, O.~T., and Colonius, T., \enquote{Spectral proper
  orthogonal decomposition and its relationship to dynamic mode decomposition
  and resolvent analysis,} \emph{Journal of Fluid Mechanics}, Vol. 847, 2018,
  pp. 821--867.

\bibitem[{Ren et~al.(2021)Ren, Mao, and Fu}]{ren2021image}
Ren, J., Mao, X., and Fu, S., \enquote{Image-based flow decomposition using
  empirical wavelet transform,} \emph{Journal of Fluid Mechanics}, Vol. 906,
  2021, p. A22.

\bibitem[{Floryan and Graham(2021)}]{floryan2021discovering}
Floryan, D., and Graham, M.~D., \enquote{Discovering multiscale and
  self-similar structure with data-driven wavelets,} \emph{Proceedings of the
  National Academy of Sciences}, Vol. 118, No.~1, 2021, p. e2021299118.

\bibitem[{Schmidt and Schmid(2019)}]{schmidt2019conditional}
Schmidt, O.~T., and Schmid, P.~J., \enquote{A conditional space--time POD
  formalism for intermittent and rare events: example of acoustic bursts in
  turbulent jets,} \emph{Journal of Fluid Mechanics}, Vol. 867, 2019.

\bibitem[{Frame and Towne(2022)}]{frame2022space}
Frame, P., and Towne, A., \enquote{Space-time POD and the Hankel matrix,}
  \emph{arXiv preprint arXiv:2206.08995}, 2022.

\bibitem[{Henningson et~al.(1993)Henningson, Lundbladh, and
  Johansson}]{henningson1993mechanism}
Henningson, D.~S., Lundbladh, A., and Johansson, A.~V., \enquote{A mechanism
  for bypass transition from localized disturbances in wall-bounded shear
  flows,} \emph{Journal of Fluid Mechanics}, Vol. 250, 1993, pp. 169--207.

\bibitem[{Cand{\`e}s and Wakin(2008)}]{candes2008introduction}
Cand{\`e}s, E.~J., and Wakin, M.~B., \enquote{An introduction to compressive
  sampling,} \emph{IEEE signal processing magazine}, Vol.~25, No.~2, 2008, pp.
  21--30.

\bibitem[{Brunton et~al.(2016)Brunton, Proctor, and Kutz}]{brunton2016sindy}
Brunton, S.~L., Proctor, J.~L., and Kutz, J.~N., \enquote{Discovering governing
  equations from data by sparse identification of nonlinear dynamical systems,}
  \emph{Proceedings of the National Academy of Sciences}, Vol. 113, No.~15,
  2016, pp. 3932--3937.

\bibitem[{Loiseau and Brunton(2018)}]{loiseau2018constrained}
Loiseau, J.-C., and Brunton, S.~L., \enquote{Constrained sparse Galerkin
  regression,} \emph{Journal of Fluid Mechanics}, Vol. 838, 2018, pp. 42--67.

\bibitem[{Rubini et~al.(2020)Rubini, Lasagna, and Da~Ronch}]{rubini2020l1}
Rubini, R., Lasagna, D., and Da~Ronch, A., \enquote{The l1-based sparsification
  of energy interactions in unsteady lid-driven cavity flow,} \emph{Journal of
  Fluid Mechanics}, Vol. 905, 2020.

\bibitem[{Jovanovi{\'c} et~al.(2014)Jovanovi{\'c}, Schmid, and
  Nichols}]{jovanovic2014dmdsp}
Jovanovi{\'c}, M.~R., Schmid, P.~J., and Nichols, J.~W.,
  \enquote{Sparsity-promoting dynamic mode decomposition,} \emph{Physics of
  Fluids (1994-present)}, Vol.~26, No.~2, 2014, 024103.

\bibitem[{Tu et~al.(2014)Tu, Rowley, Kutz, and Shang}]{tu2014compressed}
Tu, J.~H., Rowley, C.~W., Kutz, J.~N., and Shang, J.~K., \enquote{Spectral
  analysis of fluid flows using sub-{Nyquist}-rate {PIV} data,}
  \emph{Experiments in Fluids}, Vol.~55, No.~9, 2014, pp. 1--13.

\bibitem[{Skene et~al.(2022)Skene, Yeh, Schmid, and
  Taira}]{skene2022sparseForcing}
Skene, C., Yeh, C.-A., Schmid, P., and Taira, K., \enquote{Sparsifying the
  Resolvent Forcing Mode via Gradient-Based Optimisation,} \emph{Journal of
  Fluid Mechanics}, Vol. 944, 2022.

\bibitem[{Sharma and McKeon(2013)}]{sharma2013resolvent}
Sharma, A.~S., and McKeon, B.~J., \enquote{On coherent structure in wall
  turbulence,} \emph{Journal of Fluid Mechanics}, Vol. 728, 2013, pp. 196--238.

\bibitem[{Luhar et~al.(2015)Luhar, Sharma, and McKeon}]{luhar2015compliant}
Luhar, M., Sharma, A.~S., and McKeon, B.~J., \enquote{A framework for studying
  the effect of compliant surfaces on wall turbulence,} \emph{Journal of Fluid
  Mechanics}, Vol. 768, 2015, pp. 415--441.

\bibitem[{McKeon(2017)}]{mckeon2017engine}
McKeon, B.~J., \enquote{The engine behind (wall) turbulence: perspectives on
  scale interactions,} \emph{Journal of Fluid Mechanics}, Vol. 817, 2017,
  p.~P1.

\bibitem[{Dawson and McKeon(2019)}]{dawson2019shape}
Dawson, S. T.~M., and McKeon, B.~J., \enquote{On the shape of resolvent modes
  in wall-bounded turbulence,} \emph{Journal of Fluid Mechanics}, Vol. 877,
  2019, pp. 682--716.

\bibitem[{Lesshafft et~al.(2019)Lesshafft, Semeraro, Jaunet, Cavalieri, and
  Jordan}]{lesshafft2019resolvent}
Lesshafft, L., Semeraro, O., Jaunet, V., Cavalieri, A. V.~G., and Jordan, P.,
  \enquote{Resolvent-based modeling of coherent wave packets in a turbulent
  jet,} \emph{Physical Review Fluids}, Vol.~4, No.~6, 2019, p. 063901.

\bibitem[{Abreu et~al.(2020)Abreu, Cavalieri, Schlatter, Vinuesa, and
  Henningson}]{abreu2020spectral}
Abreu, L.~I., Cavalieri, A.~V., Schlatter, P., Vinuesa, R., and Henningson,
  D.~S., \enquote{Spectral proper orthogonal decomposition and resolvent
  analysis of near-wall coherent structures in turbulent pipe flows,}
  \emph{Journal of Fluid Mechanics}, Vol. 900, 2020.

\bibitem[{Towne et~al.(2020)Towne, Lozano-Dur{\'a}n, and
  Yang}]{towne2020resolvent}
Towne, A., Lozano-Dur{\'a}n, A., and Yang, X., \enquote{Resolvent-based
  estimation of space--time flow statistics,} \emph{Journal of Fluid
  Mechanics}, Vol. 883, 2020.

\bibitem[{Pickering et~al.(2021)Pickering, Towne, Jordan, and
  Colonius}]{pickering2021resolvent}
Pickering, E., Towne, A., Jordan, P., and Colonius, T.,
  \enquote{Resolvent-based modeling of turbulent jet noise,} \emph{The Journal
  of the Acoustical Society of America}, Vol. 150, No.~4, 2021, pp. 2421--2433.

\bibitem[{Yeh et~al.(2020)Yeh, Benton, Taira, and Garmann}]{yeh2020resolvent}
Yeh, C.-A., Benton, S.~I., Taira, K., and Garmann, D.~J., \enquote{Resolvent
  analysis of an airfoil laminar separation bubble at Re= 500 000,}
  \emph{Physical Review Fluids}, Vol.~5, No.~8, 2020, p. 083906.

\bibitem[{Bae et~al.(2020)Bae, Dawson, and McKeon}]{bae2020resolvent}
Bae, H.~J., Dawson, S. T.~M., and McKeon, B.~J., \enquote{Resolvent-based study
  of compressibility effects on supersonic turbulent boundary layers,}
  \emph{Journal of Fluid Mechanics}, Vol. 883, 2020, p. A29.

\bibitem[{Padovan et~al.(2020)Padovan, Otto, and Rowley}]{harmonic2020padovan}
Padovan, A., Otto, S., and Rowley, C., \enquote{Analysis of amplification
  mechanisms and cross-frequency interactions in nonlinear flows via the
  harmonic resolvent,} \emph{Journal of Fluid Mechanics}, Vol. 900, 2020.

\bibitem[{Padovan and Rowley(2022)}]{padovan2022analysis}
Padovan, A., and Rowley, C.~W., \enquote{Analysis of the dynamics of
  subharmonic flow structures via the harmonic resolvent: Application to vortex
  pairing in an axisymmetric jet,} \emph{Physical Review Fluids}, Vol.~7,
  No.~7, 2022, p. 073903.

\bibitem[{Hein and B{\"u}hler(2010)}]{hein2010inverse}
Hein, M., and B{\"u}hler, T., \enquote{An inverse power method for nonlinear
  eigenproblems with applications in 1-spectral clustering and sparse PCA,}
  \emph{Advances in Neural Information Processing Systems}, Vol.~23, 2010.

\bibitem[{Jolliffe et~al.(2003)Jolliffe, Trendafilov, and
  Uddin}]{jolliffe2003modified}
Jolliffe, I.~T., Trendafilov, N.~T., and Uddin, M., \enquote{A modified
  principal component technique based on the LASSO,} \emph{Journal of
  computational and Graphical Statistics}, Vol.~12, No.~3, 2003, pp. 531--547.

\bibitem[{Zou et~al.(2006)Zou, Hastie, and Tibshirani}]{zou2006sparse}
Zou, H., Hastie, T., and Tibshirani, R., \enquote{Sparse principal component
  analysis,} \emph{Journal of computational and graphical statistics}, Vol.~15,
  No.~2, 2006, pp. 265--286.

\bibitem[{Sigg and Buhmann(2008)}]{sigg2008expectation}
Sigg, C.~D., and Buhmann, J.~M., \enquote{Expectation-maximization for sparse
  and non-negative PCA,} \emph{Proceedings of the 25th international conference
  on Machine learning}, 2008, pp. 960--967.

\bibitem[{Journ{\'e}e et~al.(2010)Journ{\'e}e, Nesterov, Richt{\'a}rik, and
  Sepulchre}]{journee2010generalized}
Journ{\'e}e, M., Nesterov, Y., Richt{\'a}rik, P., and Sepulchre, R.,
  \enquote{Generalized power method for sparse principal component analysis.}
  \emph{Journal of Machine Learning Research}, Vol.~11, No.~2, 2010.

\bibitem[{Zou and Xue(2018)}]{zou2018selective}
Zou, H., and Xue, L., \enquote{A selective overview of sparse principal
  component analysis,} \emph{Proceedings of the IEEE}, Vol. 106, No.~8, 2018,
  pp. 1311--1320.

\bibitem[{B{\"u}hler(2014)}]{buhler2014flexible}
B{\"u}hler, T., \enquote{A flexible framework for solving constrained ratio
  problems in machine learning,} Ph.D. thesis, Saarland University, 2014.

\bibitem[{Bae et~al.(2018)Bae, Lozano-Duran, Bose, and
  Moin}]{bae2018turbulence}
Bae, H.~J., Lozano-Duran, A., Bose, S., and Moin, P., \enquote{Turbulence
  intensities in large-eddy simulation of wall-bounded flows,} \emph{Physical
  Review Fluids}, Vol.~3, No.~1, 2018, p. 014610.

\bibitem[{Bae et~al.(2019)Bae, Lozano-Dur{\'a}n, Bose, and
  Moin}]{bae2019dynamic}
Bae, H.~J., Lozano-Dur{\'a}n, A., Bose, S.~T., and Moin, P., \enquote{Dynamic
  slip wall model for large-eddy simulation,} \emph{Journal of fluid
  mechanics}, Vol. 859, 2019, pp. 400--432.

\bibitem[{Lozano-Dur{\'a}n and Bae(2019)}]{lozano2019characteristic}
Lozano-Dur{\'a}n, A., and Bae, H.~J., \enquote{Characteristic scales of
  Townsend’s wall-attached eddies,} \emph{Journal of fluid mechanics}, Vol.
  868, 2019, pp. 698--725.

\bibitem[{Orlandi(2000)}]{orlandi2000fluid}
Orlandi, P., \emph{Fluid flow phenomena: a numerical toolkit}, Vol.~55,
  Springer Science \& Business Media, 2000.

\bibitem[{Kim and Moin(1985)}]{kim1985application}
Kim, J., and Moin, P., \enquote{Application of a fractional-step method to
  incompressible Navier-Stokes equations,} \emph{Journal of computational
  physics}, Vol.~59, No.~2, 1985, pp. 308--323.

\bibitem[{Wray(1990)}]{wray1990minimal}
Wray, A.~A., \enquote{Minimal storage time advancement schemes for spectral
  methods,} \emph{NASA Ames Research Center, California, Report No. MS}, Vol.
  202, 1990.

\bibitem[{Weideman and Reddy(2000)}]{weideman2000matlab}
Weideman, J.~A., and Reddy, S.~C., \enquote{A {M}ATLAB differentiation matrix
  suite,} \emph{ACM Transactions on Mathematical Software (TOMS)}, Vol.~26,
  No.~4, 2000, pp. 465--519.

\bibitem[{Jim{\'e}nez and Pinelli(1999)}]{jimenez1999autonomous}
Jim{\'e}nez, J., and Pinelli, A., \enquote{The autonomous cycle of near-wall
  turbulence,} \emph{Journal of Fluid Mechanics}, Vol. 389, 1999, pp. 335--359.

\bibitem[{Landahl(1975)}]{landahl1975wave}
Landahl, M.~T., \enquote{Wave breakdown and turbulence,} \emph{SIAM Journal on
  Applied Mathematics}, Vol.~28, No.~4, 1975, pp. 735--756.

\bibitem[{Landahl(1980)}]{landahl1980note}
Landahl, M.~T., \enquote{A note on an algebraic instability of inviscid
  parallel shear flows,} \emph{Journal of Fluid Mechanics}, Vol.~98, No.~2,
  1980, pp. 243--251.

\bibitem[{Orr(1907)}]{orr1907stability}
Orr, W.~M., \enquote{The stability or instability of the steady motions of a
  perfect liquid and of a viscous liquid. {P}art {II}: A viscous liquid,}
  \emph{Proceedings of the Royal Irish Academy. Section A: Mathematical and
  Physical Sciences}, JSTOR, 1907, pp. 69--138.

\bibitem[{Jim{\'e}nez(2013)}]{jimenez2013linear}
Jim{\'e}nez, J., \enquote{How linear is wall-bounded turbulence?} \emph{Physics
  of Fluids}, Vol.~25, No.~11, 2013, p. 110814.

\bibitem[{Dawson and McKeon(2020)}]{dawson2020ijhff}
Dawson, S. T.~M., and McKeon, B.~J., \enquote{Prediction of resolvent mode
  shapes in supersonic turbulent boundary layers,} \emph{International Journal
  of Heat and Fluid Flow}, Vol.~85, 2020, p. 108677.

\bibitem[{Akhavan et~al.(1991)Akhavan, Kamm, and
  Shapiro}]{akhavan1991investigation}
Akhavan, R., Kamm, R., and Shapiro, A., \enquote{An investigation of transition
  to turbulence in bounded oscillatory Stokes flows Part 1. Experiments,}
  \emph{Journal of Fluid Mechanics}, Vol. 225, 1991, pp. 395--422.

\bibitem[{Verzicco and Vittori(1996)}]{verzicco1996direct}
Verzicco, R., and Vittori, G., \enquote{Direct simulation of transition in
  Stokes boundary layers,} \emph{Physics of Fluids}, Vol.~8, No.~6, 1996, pp.
  1341--1343.

\bibitem[{Vittori and Verzicco(1998)}]{vittori1998direct}
Vittori, G., and Verzicco, R., \enquote{Direct simulation of transition in an
  oscillatory boundary layer,} \emph{Journal of Fluid Mechanics}, Vol. 371,
  1998, pp. 207--232.

\bibitem[{Costamagna et~al.(2003)Costamagna, Vittori, and
  Blondeaux}]{costamagna2003coherent}
Costamagna, P., Vittori, G., and Blondeaux, P., \enquote{Coherent structures in
  oscillatory boundary layers,} \emph{Journal of Fluid Mechanics}, Vol. 474,
  2003, pp. 1--33.

\end{thebibliography}

\end{document}